\PassOptionsToPackage{dvipsnames}{xcolor}

\documentclass[authoryear,final,3p,times,twocolumn]{elsarticle}

\usepackage{xcolor}
\usepackage{colortbl}

\usepackage{amsmath,amssymb,amsfonts,mathtools}
\usepackage{amsthm}

\usepackage{graphicx}
\usepackage{booktabs}
\usepackage{multirow}
\usepackage{makecell}
\usepackage{array}
\usepackage{tabularx}
\usepackage{arydshln}
\usepackage{adjustbox}
\usepackage{textcomp}
\usepackage{enumitem}

\usepackage{pgfplots}
\pgfplotsset{compat=1.18}

\usepackage[linesnumbered,ruled,vlined]{algorithm2e}
\SetKwInput{KwInit}{Init}

\usepackage{float}
\usepackage{placeins}
\usepackage{stfloats}

\usepackage{subfig}

\usepackage[most]{tcolorbox}

\usepackage{tikz}
\usetikzlibrary{
    arrows.meta,
    positioning,
    fit,
    calc,
    backgrounds,
    shapes.geometric
}

\usepackage{listings}

\usepackage{xurl}
\usepackage{url}

\usepackage[colorlinks=true,
    linkcolor=MidnightBlue,
    citecolor=ForestGreen,
    urlcolor=BrickRed
]{hyperref}

\usepackage{pifont}

\newcommand{\method}{\textsc{SPECTRA-Siam}}
\newcommand{\concat}{\mathbin\Vert}
\newcommand{\relu}[1]{\left[#1\right]_{+}}

\newcommand{\safeincludegraphics}[2][]{%
  \IfFileExists{#2}{%
    \includegraphics[#1]{#2}%
  }{%
    \fbox{\parbox[c][3cm][c]{0.90\linewidth}{%
      \centering Missing draft figure:\\[-1mm]
      \texttt{\detokenize{#2}}%
    }}%
  }%
}

\theoremstyle{plain}

\definecolor{titlegray}{HTML}{555555}
\definecolor{framegray}{gray}{0.6}
\definecolor{lightgray}{gray}{0.97}

\newtcolorbox{promptbox}[1][]{
    enhanced,
    colback=lightgray,
    colframe=framegray,
    coltitle=white,
    colbacktitle=titlegray,
    fonttitle=\bfseries\itshape\normalsize,
    title=#1,
    attach boxed title to top left={xshift=0.6cm,yshift=-3mm},
    boxed title style={
        boxrule=0.8pt,
        colframe=titlegray,
        top=3pt,
        bottom=3pt,
        left=8pt,
        right=8pt,
        rounded corners=south,
    },
    width=0.90\columnwidth,
    before skip=6pt,
    after skip=12pt,
    boxrule=0.8pt,
    sharp corners=south,
    rounded corners,
    top=8pt,
    bottom=6pt,
    left=8pt,
    right=8pt,
    fontupper=\normalsize,
    coltext=black,
}

\lstdefinelanguage{diff}{
  morecomment=[f][\color{gray}]{@@},
  morecomment=[f][\color{gray}]{---},
  morecomment=[f][\color{gray}]{+++},
  morecomment=[f][\color{red}]{-},
  morecomment=[f][\color{green!60!black}]{+},
}
\lstdefinestyle{diff}{
  language=diff,
  basicstyle=\ttfamily\footnotesize,
  columns=fullflexible,
  keepspaces=true,
  showstringspaces=false,
  breaklines=true
}
\definecolor{deepblue}{RGB}{0, 0, 139} 

\newtcolorbox{boxnote}{
  colback=white,
  boxrule=0.4pt,
  borderline west={2pt}{0pt}{black!60},
}

\newcommand{\gain}[1]{\textcolor{green!50!black}{(+#1$\uparrow$)}}

\newcommand{\smallgain}[1]{\textcolor{gray}{(+#1$\uparrow$)}}
\newcommand{\smalldrop}[1]{\textcolor{gray}{(-#1$\downarrow$)}}
\newcommand{\samev}{\textcolor{gray}{(.00)}}

\newcommand{\subfigref}[2]{%
    \hyperref[#2]{Figure~\ref*{#1}\subref*{#2}}%
}

\begin{document}

\begin{frontmatter}



\title{Learning Spectral Representations of Code through Latent Graph Learning for Generalizable Cross-Language Code Clone Detection}


\author[mce]{Mohsen Hesamolhokama}
\ead{hokama@ce.sharif.edu}

\author[math]{Ali Sadeghi}
\ead{ali.sadeghi@sharif.edu}

\author[math]{Kousha Moeini}
\ead{Kousha.m83@sharif.edu}

\author[math]{Behnam Rohani}
\ead{behnam.rohani058@sharif.edu}

\author[mce,cor1]{Mohammadamin Fazli}
\ead{fazli@sharif.edu}

\author[mce]{Jafar Habibi}
\ead{jhabibi@sharif.edu}

\address[mce]{Department of Computer Engineering, Sharif University of Technology, Tehran, Iran}
\address[math]{Department of Mathematical Sciences, Sharif University of Technology, Tehran, Iran}

\cortext[cor1]{Corresponding author.}

\begin{abstract}
Current code clone detection (CCD) methods rely on fixed, language-specific graph representations like abstract syntax trees (ASTs) or program dependency graphs (PDGs). Because functionally identical code fragments can yield wildly different structures, these rigid graphs produce non-discriminative spectra that perform close to chance. To address this, we propose \method{}, a Siamese latent graph learning network that learns a latent space such that the graph's spectrum serves as a discriminative signature of code functionality by optimizing downstream CCD performance. Given a fragment's AST and data-dependencies, \method{} induces a fixed-size weighted latent graph through soft slot assignment and multi-head attention, and extracts a multi-scale spectral representation from its normalized Laplacian. Mapping all fragments into this shared space yields comparable spectra across programming languages. Experiments on BigCloneBench, AtCoder, and a four-language CodeNet benchmark (Java, Python, C++, C\#) support this design choice. Using the same downstream classifier, moving from fixed to learned latent graphs spectra jumps F1 from 0.37 to 0.67 on BigCloneBench and accuracy from 0.60 to 0.71 on AtCoder. On CodeNet, the full model reaches 0.69 accuracy in four epochs and 0.79 after thirty epochs. In bridge-assisted language transfer across 60 unseen paths, \method{}'s performance degrades by only 0.058, versus 0.112--0.228 for baselines, showing that learned graph spectra provide a highly generalizable representation for cross-language clone detection.
\end{abstract}



\end{frontmatter}

\section{Introduction}
\label{sec:introduction}
Nearly every automated software engineering task begins with source code, and its effectiveness depends on how the underlying program is represented \citep{allamanis2018survey}. A useful representation should preserve the structural and semantic information required for automated reasoning about programs. This requirement is central to tasks such as code search \citep{gu2018deepcs}, plagiarism detection \citep{schleimer2003winnowing}, and vulnerability analysis \citep{zhou2019devign}. Among these tasks, Code Clone Detection (CCD) aims to identify fragments that are identical, syntactically similar, or functionally equivalent. Such clones are prevalent in real-world software systems and increase maintenance costs because modifications to one fragment may need to be consistently propagated to its counterparts \citep{juergens2009clones,roy2007survey}.

The effectiveness of CCD depends strongly on whether a representation captures meaningful similarities beyond surface-level patterns. Early approaches represented code primarily as text or token sequences \citep{schleimer2003winnowing}. Later methods incorporated abstract syntax trees (ASTs) using tree-based neural networks \citep{mou2016tbcnn}, while another line learned distributed representations from code tokens and their contexts \citep{mikolov2013word2vec,alon2019code2vec}. More recently, the Transformer architecture \citep{vaswani2017attention} and pre-trained models such as BERT \citep{devlin2019bert} have been adapted to source code, resulting in models such as CodeBERT \citep{feng2020codebert} and GraphCodeBERT \citep{guo2021graphcodebert}. All these approaches share the goal of mapping semantically related fragments to similar representations. Contrastive learning further strengthens this objective by encouraging such alignment through semantics-preserving transformations \citep{chen2020simclr,jain2021contracode,bui2021corder} or self-supervised objectives derived from AST structure \citep{bui2021infercode}.

Despite these advances, capturing program semantics remains a major challenge for CCD. Clones are commonly categorized into four types \citep{roy2007survey}; while the first three involve varying degrees of syntactic similarity, Type-IV clones represent functionally equivalent fragments and are difficult to detect \citep{alomari2020semanticclonebench}. Existing CCD approaches have explored learned representations through deep feature models \citep{white2016deep,wei2017cdlh}, AST-based encoders \citep{zhang2019astnn}, and graph-based representations that incorporate richer program structures \citep{zhao2018deepsim,wang2020faast,wu2020scdetector}. However, semantic clone detectors often generalize poorly to projects outside their training data, limiting their ability to handle unseen code \citep{sonnekalb2022generalizability}. Graph-based methods incorporate richer program information through ASTs, CFGs, DDGs, or CPGs \citep{yamaguchi2014cpg,zhao2018deepsim,wang2020faast,wu2020scdetector}, but these structures are fixed before training as the model learns how to encode a given graph. This limitation becomes significant in cross-language clone detection, where equivalent functionality may be expressed through different syntactic structures and language constructs \citep{perez2019crosslang,zhang2024fsdclcd,alam2023gptclonebench}.

Graph spectra offer an alternative by summarizing graph structure through the eigenvalues of adjacency or Laplacian matrices. These eigenvalues are invariant to node relabeling \citep{chung1997spectral,mohar1991laplacian}, capture global connectivity properties \citep{fiedler1973algebraic}, and can be truncated to a bounded number of leading components \citep{defferrard2016cnngraph}. Although spectral descriptors have been widely used as structural signatures in geometry processing \citep{sun2009hks,zhang2010spectralmesh} and graph mining \citep{belkin2003laplacian,narayanan2017graph2vec}, their use in software engineering has largely focused on software architecture analysis, such as partitioning module dependency graphs \citep{mitchell2006software}. To our knowledge, the graph spectrum has not been explored as a representation of individual program fragments for source code representation learning or clone detection.

This motivates our central design choice: we do not commit to any predefined program graph. We learn a graph whose spectral properties are optimized for distinguishing functional equivalence. This perspective follows latent graph learning, where the graph structure itself is treated as a learnable variable jointly optimized with the prediction task \citep{kipf2018nri,franceschi2019lds,chen2020idgl,zhu2021gslsurvey,saha2023adaptive}. We propose \method{}, a Siamese model that transforms the AST and projected data-dependence information of each code fragment into a language-independent latent graph. The learned graph provides a compact and comparable structure across fragments of different sizes, while its spectral properties serve as the basis for the final representation. Unlike approaches that commit to a fixed AST, CFG, DDG, or CPG structure \citep{yamaguchi2014cpg}, \method{} lets clone supervision determine the topology that best captures functional similarity. A shared representation space across parsers further enables cross-language clone detection. 

\definecolor{kwcolor}{RGB}{0,0,180}
\definecolor{cmtcolor}{RGB}{0,128,0}
\definecolor{strcolor}{RGB}{170,0,0}
\definecolor{bgcolor}{RGB}{248,248,248}
\lstdefinestyle{javastyle}{
  language=Java, basicstyle=\scriptsize\ttfamily,
  keywordstyle=\color{kwcolor}\bfseries, commentstyle=\color{cmtcolor}\itshape,
  stringstyle=\color{strcolor}, numberstyle=\tiny\color{gray},
  backgroundcolor=\color{bgcolor}, frame=single, framesep=2pt,
  showstringspaces=false, breaklines=true, columns=fullflexible, keepspaces=true,
  aboveskip=2pt, belowskip=2pt,
}

\subsection{A Motivating Example}
\label{sec:motivating_exp}

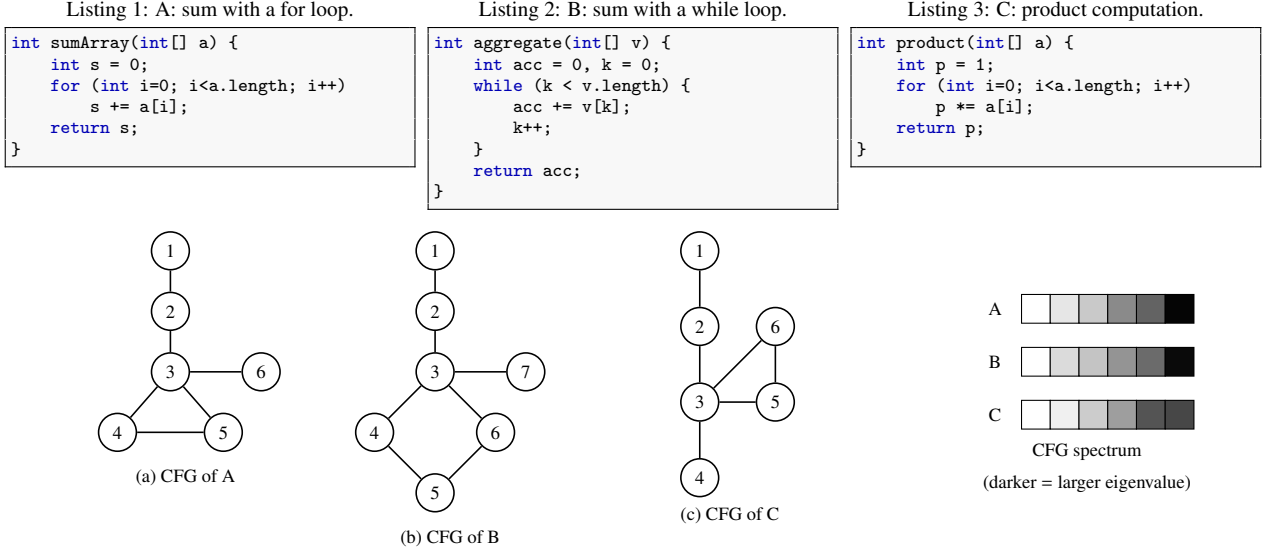
\begin{figure*}[t]
\centering

\begin{minipage}[t]{0.32\linewidth}
\begin{lstlisting}[style=javastyle,caption={A: sum with a for loop.},label={lst:for}]
int sumArray(int[] a) {
    int s = 0;
    for (int i=0; i<a.length; i++)
        s += a[i];
    return s;
}
\end{lstlisting}
\end{minipage}\hfill
\begin{minipage}[t]{0.32\linewidth}
\begin{lstlisting}[style=javastyle,caption={B: sum with a while loop.},label={lst:while}]
int aggregate(int[] v) {
    int acc = 0, k = 0;
    while (k < v.length) {
        acc += v[k];
        k++;
    }
    return acc;
}
\end{lstlisting}
\end{minipage}\hfill
\begin{minipage}[t]{0.32\linewidth}
\begin{lstlisting}[style=javastyle,caption={C: product computation.},label={lst:product}]
int product(int[] a) {
    int p = 1;
    for (int i=0; i<a.length; i++)
        p *= a[i];
    return p;
}
\end{lstlisting}
\end{minipage}

\vspace{5pt}

\begin{minipage}[c]{0.70\linewidth}
\centering
\begin{tikzpicture}[
vtx/.style={
circle,
draw,
semithick,
fill=white,
minimum size=5mm,
inner sep=0pt,
font=\scriptsize
},
every path/.style={semithick}
]

\node[vtx](a1) at (0,0){1};
\node[vtx](a2) at (0,-0.8){2};
\node[vtx](a3) at (0,-1.6){3};
\node[vtx](a4) at (-0.7,-2.4){4};
\node[vtx](a5) at (0.7,-2.4){5};
\node[vtx](a6) at (1.2,-1.6){6};

\draw(a1)--(a2)--(a3);
\draw(a3)--(a4)--(a5)--(a3);
\draw(a3)--(a6);

\node[font=\scriptsize] at (0.2,-3.0){(a) CFG of A};

\begin{scope}[xshift=35mm]

\node[vtx](b1) at (0,0){1};
\node[vtx](b2) at (0,-0.8){2};
\node[vtx](b3) at (0,-1.6){3};
\node[vtx](b4) at (-0.8,-2.4){4};
\node[vtx](b5) at (0,-3.2){5};
\node[vtx](b6) at (0.8,-2.4){6};
\node[vtx](b7) at (1.2,-1.6){7};

\draw(b1)--(b2)--(b3);
\draw(b3)--(b4)--(b5)--(b6)--(b3);
\draw(b3)--(b7);

\node[font=\scriptsize] at (0.2,-3.8){(b) CFG of B};

\end{scope}

\begin{scope}[xshift=70mm]

\node[vtx](c1) at (0,0){1};
\node[vtx](c2) at (0,-1){2};
\node[vtx](c3) at (0,-2){3};
\node[vtx](c4) at (0,-3){4};
\node[vtx](c5) at (1,-2){5};
\node[vtx](c6) at (1,-1){6};

\draw(c1)--(c2)--(c3);
\draw(c3)--(c4);
\draw(c3)--(c5)--(c6)--(c3);

\node[font=\scriptsize] at (0.4,-3.5){(c) CFG of C};

\end{scope}

\end{tikzpicture}
\end{minipage}
\hfill
\begin{minipage}[c]{0.27\linewidth}
\centering

\begin{tikzpicture}[every path/.style={thin}]

\def\cw{0.38}

\node[font=\scriptsize] at (-0.35,0){A};

\foreach \i/\p in {0/0,1/10,2/21,3/46,4/61,5/98}
{
\fill[black!\p] (\i*\cw,-0.19)
rectangle ++(\cw,0.38);
\draw(\i*\cw,-0.19)
rectangle ++(\cw,0.38);
}

\node[font=\scriptsize] at (-0.35,-0.7){B};

\foreach \i/\p in {0/0,1/14,2/23,3/42,4/58,5/96}
{
\fill[black!\p] (\i*\cw,-0.89)
rectangle ++(\cw,0.38);
\draw(\i*\cw,-0.89)
rectangle ++(\cw,0.38);
}

\node[font=\scriptsize] at (-0.35,-1.4){C};

\foreach \i/\p in {0/0,1/6,2/20,3/38,4/67,5/72}
{
\fill[black!\p] (\i*\cw,-1.59)
rectangle ++(\cw,0.38);
\draw(\i*\cw,-1.59)
rectangle ++(\cw,0.38);
}

\end{tikzpicture}

\vspace{2pt}

{\scriptsize
CFG spectrum\\
(darker = larger eigenvalue)}

\end{minipage}

\caption{Fixed CFG spectra mislead clone detection. The semantically equivalent pair (A, B) is spectrally farther apart
(distance 2.14) than either is from the unrelated function C (1.02 and 1.24).}

\label{fig:motivating_exp}
\end{figure*}

\autoref{fig:motivating_exp} illustrates the limitation of fixed CFG for clone detection. Functions A and B implement the same array-summation behavior, but different loop structures and intermediate computations produce different CFGs. Their Laplacian spectra, $\{0,\,0.486,\,1.000,\,2.428,\,3.000,\,5.086\}$ and $\{0,\,0.433,\allowbreak\,0.851,\,2.000,\,2.302,\,3.113\}$, have a distance of $2.14$. The unrelated product function C has smaller distances from A and B, with distances of $1.02$ and $1.24$, respectively. A more suitable representation should therefore learn graph structures that are more aligned with semantic similarity of code.

\subsection{Contributions}
\label{subsec:contributions}
The main contributions of this paper are as follows:

\begin{itemize}
    \item \textbf{A Novel Latent Graph Learning Network for Spectral Code Representation.} We propose \method{}, a Siamese latent graph learning network with soft slot assignment and multi-head attention that learns a fixed-size weighted latent graph and derives a multi-scale spectral representation from its normalized Laplacian. Even without its original prediction head, \method{} significantly outperforms fixed-graph spectral baselines (AST, CFG, DDG, and CPG) under identical downstream classifiers, improving spectral F1 from .37 to .67 on BigCloneBench and from .65 to .73 on AtCoder (\autoref{tab:rq1}).

    \item \textbf{Cross-Language Robustness under Language Shifts and Transfer Paths.} We evaluate cross-language generalization using bridge-assisted transfer paths, where a model learns from one or more non-target language pairs and is tested on an unseen target language pair. Across 60 such transfer paths (\autoref{tab:rq3_effects}), \method{} shows the smallest average degradation under language shift (.058), compared with .112--.228 for RtvNN, DeepSim, ASTNN, and GraphCodeBERT . \method{} is also the only method above chance on all six cross-language pairs without cross-language supervision (\autoref{tab:rq3_controls}).
\end{itemize}

\subsection{Paper Organization} 
\label{subsec:outline} 
The rest of this paper is organized as follows. \autoref{sec:related_work} reviews related work. \autoref{sec:methodology} presents our latent graph induction framework and spectral representation. \autoref{sec:experimental_design} describes the datasets, baselines, and evaluation protocols, while \autoref{sec:exp_results} reports the evaluation results on single-language and cross-language benchmarks. \autoref{sec:discussion} analyzes the learned spectral representation space, and \autoref{sec:conclusion} summarizes our findings and concludes the paper.

\section{Related Work}
\label{sec:related_work}

\subsection{Code Clone Detection}
\label{subsec:ccd}
Code clone detection has been studied for more than two decades. Methods are often distinguished by the representation used to compare code fragments. Roy and Cordy \citep{roy2007survey} classify clones into four types. Type-1 clones differ only in formatting or naming. Type-2 and Type-3 clones allow increasingly larger syntactic changes. Type-4 clones implement the same functionality despite having different syntax.

Early scalable detectors relied mainly on text or tokens. \emph{CCFinder} \citep{kamiya2002ccfinder} normalizes token sequences before comparison. \citet{ducasse1999language} instead compare source code at the line level in a language-independent manner, while \emph{CP-Miner} \citep{li2006cpminer} mines copy-paste patterns in large software systems. \emph{NiCad} \citep{roy2008nicad} combines source transformation, pretty-printing, normalization, and line-based comparison to find near-miss clones. Later methods improved scalability through more selective comparison. \emph{SourcererCC} \citep{sajnani2016sourcerercc} uses an inverted index, \emph{CCAligner} \citep{wang2018ccaligner} compares overlapping code windows, and \emph{Oreo} \citep{saini2018oreo} uses filtering before clone comparison. \emph{StoneDetector} \citep{heinze2025stonedetector} takes a different route by extracting paths from dominator trees and comparing their encodings. These methods can scale to large codebases, but text and token similarity alone becomes less reliable as two implementations diverge syntactically.

Tree-based methods make syntax explicit. \citet{baxter1998clone} detect clones by matching AST subtrees, while \emph{DECKARD} \citep{jiang2007deckard} maps AST subtrees to numerical vectors and clusters similar vectors. ASTs provide more information than raw text, but syntax alone may still be insufficient when equivalent code is reorganized substantially. Program graphs can capture relations that are less tied to source order. Komondoor and Horwitz \citep{komondoor2001slicing} and Krinke \citep{krinke2001identifying} use program dependence graphs (PDGs), whose edges describe control and data dependencies. Such representations can reveal similarities that are difficult to see from tokens or syntax trees, although graph and subgraph matching can be computationally expensive.

More recent approaches learn code representations from data. \citet{white2016deep} were among the early works to apply deep learning to clone detection using lexical and syntactic information. \emph{CDLH} \citep{wei2017cdlh} formulates functional clone detection as supervised hashing over AST-based LSTMs. \emph{ASTNN} \citep{zhang2019astnn} splits an AST into statement-level subtrees and encodes their sequence with recurrent neural networks (RNNs). Other approaches use tree-based convolution \citep{yu2019neural} or recursive neural models over ASTs \citep{buch2019learning}. Another line learns from engineered structural features. \citet{sheneamer2018detection} combine features from bytecode dependency graphs, PDGs, and ASTs to detect semantic clones and obfuscated code, \citet{sheneamer2021multiple} blend several similarity measures for consensus-driven classification, and \citet{sudhamani2019code} compare control-statement and program-level metrics directly.

GNN-based methods incorporate additional program relations into the learned representation. \citet{wang2020faast} introduce the flow-augmented AST (\emph{FA-AST}), which adds control-flow and data-flow edges to the AST and applies graph neural networks for clone detection. \citet{yu2023graphsemantics} use either a CFG or a PDG and learn graph representations with attention at several levels before comparing code fragments. For cross-language clone detection, \emph{FSD-CLCD} \citep{zhang2024fsdclcd} builds graph representations from ASTs and transfers functional information across languages through knowledge distillation. \emph{SCOTT} \citep{gao2026scott} combines a graph and a text view of each fragment with contrastive learning to cover all four clone types. Cross-language analysis has also moved beyond clone detection, for example to multilingual refactoring \citep{li2024multilingual}.

Other work focuses on computational cost or alternative input representations. \citet{zhang2023efficient} reduce Transformer cost through code token learners, \emph{CCStokener} \citep{wang2023ccstokener} enriches token representations with semantic information, and \emph{Code2Img} \citep{hu2023code2img} converts AST adjacency matrices into images processed by convolutional networks. \emph{BigCloneBench} \citep{svajlenko2014bigclonebench} is widely used to evaluate many of these learning-based clone detectors.

Large language models have recently been applied to the same task. A systematic mapping of this literature \citep{jiang2026mapping} finds that prompting improves Type-3 and Type-4 detection, but that prompt sensitivity remains a major source of variance and Type-4 clones stay difficult; broader surveys report similar limitations \citep{dou2023llmsurvey}.

\subsection{Source Code Representation}
\label{subsec:representation}
How code is represented determines which similarities a detector can observe. The AST describes the syntactic organization of a program, while the control-flow graph (CFG) describes possible execution paths and the PDG adds control and data dependencies. The code property graph (CPG) introduced by \citet{yamaguchi2014cpg} brings syntax, control flow, and data flow into a unified graph and was originally developed for vulnerability discovery.

A different line of work learns vector representations of code. \emph{code2vec} \citep{alon2019code2vec} represents a code fragment through paths sampled from its AST and combines them into a fixed-length vector. Pre-trained language models learn from much larger source-code corpora. \emph{CodeBERT} \citep{feng2020codebert} learns joint representations of programming language and natural language. \emph{GraphCodeBERT} \citep{guo2021graphcodebert} also uses data-flow information during pre-training and applies the resulting model to tasks including clone detection. Models such as ASTNN \citep{zhang2019astnn} and FA-AST \citep{wang2020faast} similarly learn representations from program trees or graphs. \citet{yu2023codeclassification} report that graph neural networks over code graphs are sensitive to how the graph is constructed, and that reported gains are hard to reproduce across settings.

\subsection{Spectral Analysis of Graphs}
\label{subsec:spectral}

Spectral graph theory studies graphs through the eigenvalues and eigenvectors of matrices associated with them, most commonly the adjacency matrix and graph Laplacian \citep{chung1997spectral,mohar1991laplacian}. The Laplacian spectrum reflects several properties of a graph. A classical example is the connection between the second-smallest Laplacian eigenvalue and graph connectivity established by Fiedler \citep{fiedler1973algebraic}. Eigenvalues are also unchanged by a permutation of the node ordering, which makes them useful when the identity or ordering of nodes should not affect the representation.

Spectral methods appear in many areas of graph analysis. Laplacian eigenvectors are used in spectral clustering \citep{shi2000normalized,ng2002spectral,vonluxburg2007tutorial} and Laplacian eigenmaps \citep{belkin2003laplacian}. Eigenvalue and eigenvector information has also been used for graph matching \citep{leordeanu2005spectral} and shape analysis \citep{sun2009hks,zhang2010spectralmesh}. In graph neural networks, \citet{bruna2014spectral} define convolution using the graph Laplacian spectrum. \citet{defferrard2016cnngraph} make these filters local and efficient through Chebyshev polynomials, and Kipf and Welling \citep{kipf2017gcn} derive a simpler graph convolution from a first-order approximation of spectral filtering.

\subsection{Latent Graph Learning}
\label{subsec:latent_graph}

Most message-passing networks assume that the graph structure is known beforehand. This assumption can be restrictive when the available graph is not a good representation for the downstream task. Latent graph learning, often studied under the broader term graph structure learning, addresses this problem by estimating the graph along with the predictive model \citep{zhu2021gslsurvey}. 

Several methods illustrate different ways to do this. Neural Relational Inference (NRI) \citep{kipf2018nri} learns a latent interaction graph from observations of a dynamical system. \citet{franceschi2019lds} learn the parameters of a graph neural network and a discrete probability distribution over graph edges through bi-level optimization. IDGL \citep{chen2020idgl} repeatedly updates node embeddings and reconstructs a graph from learned similarities between them. Other methods learn neighborhoods and allow the number of selected neighbors to vary \citep{saha2023adaptive}. Graph pooling provides another way to handle graphs of varying sizes by learning how input nodes are grouped into a smaller graph. DiffPool \citep{ying2018diffpool} learns a soft assignment from input nodes to a fixed number of clusters and produces a smaller graph at the next layer.

Latent graph learning is particularly relevant to our setting. Existing graph-based code models start from relations obtained through parsing or program analysis, such as those in an AST, CFG, PDG, CPG, or a manually augmented graph \citep{yamaguchi2014cpg,wang2020faast,yu2023graphsemantics,zhang2024fsdclcd}. Some software-engineering methods also simplify these graphs before learning. AMPLE \citep{wen2023ample}, for example, simplifies code graphs and combines information from different edge types for vulnerability detection.

\section{Methodology}
\label{sec:methodology}
Spectral representations provide compact, permutation-invariant descriptions of graph structure and have been used for graph comparison and matching \citep{wilson2008study}. Their discriminative content, however, depends on the graph being represented. Using a fixed AST, CFG, or DDG therefore commits the representation to a predefined structural view of the program. We instead learn a latent graph from semantic-clone supervision and optimize it jointly with the downstream prediction task \citep{yu2021deep}.

\subsection{Problem Formulation}
\label{sec:problem_formulation}
Let $\mathcal{L}$ denote a set of programming languages, $\mathcal{C}_{\ell}$ the set of code fragments written in language $\ell\in\mathcal{L}$, and $\mathcal{C}=\bigcup_{\ell\in\mathcal{L}}\mathcal{C}_{\ell}$. Given a labeled training set $\mathcal{D}_{\mathrm{tr}}=\{(c_i,c_j,y_{ij})\}$, where $c_i,c_j\in\mathcal{C}$,
\begin{equation}
y_{ij}=
\begin{cases}
1, & c_i\equiv_{\mathrm{func}}c_j,\\
0, & \text{otherwise},
\end{cases}
\label{eq:clone_label}
\end{equation}
indicates whether the two fragments implement the same functionality. Let $\mathcal{G}_{m,d}$ denote the space of weighted attributed graphs with $m$ latent nodes, each represented by a $d$-dimensional feature vector. We seek a parameterized code-to-graph mapping $G_{\Theta}:\mathcal{C}\rightarrow\mathcal{G}_{m,d}$ such that $G_{\Theta}(c)=(Z_{\Theta}(c),A_{\Theta}(c))$. The graph is represented by its spectral signature $s_{\Theta}(c)=\operatorname{Spec}(G_{\Theta}(c))\in\mathbb{R}^{d_s}$. Define
\begin{equation}
S_{\Theta}(c_i,c_j)
=
\cos\!\left(s_{\Theta}(c_i),s_{\Theta}(c_j)\right).
\label{eq:spectral_similarity}
\end{equation}
and let the spectral loss be
\begin{equation}
\ell_{ij}(\Theta)
=
y_{ij}(1-S_{\Theta}(c_i,c_j))
+
(1-y_{ij})[S_{\Theta}(c_i,c_j)-\mu_{\mathrm{spec}}]_{+},
\label{eq:spectral_loss}
\end{equation}
where $[x]_{+}=\max(0,x)$ and $\mu_{\mathrm{spec}}\in[0,1)$ is the negative-pair similarity margin, specifying the maximum similarity tolerated for negative pairs. The objective then is to learn $G_{\Theta}$ such that functional equivalence is preserved in the induced spectral space, i.e.,
\begin{equation}
\Theta^{*}
=
\arg\min_{\Theta}
\mathbb{E}_{(c_i,c_j,y_{ij})\sim\mathcal{D}_{\mathrm{tr}}}
\left[\ell_{ij}(\Theta)\right].
\label{eq:latent_graph_objective}
\end{equation}
At inference time, the complete model produces a clone probability $\widehat{p}_{ij}$ for each pair. Clone detection is performed as $\widehat{y}_{ij}=\mathbb{I}[S_{\Theta^{*}}(c_i,c_j)\geq\gamma]$, where $\gamma$ is the decision threshold.

\subsection{\method{}}
\label{sec:spectra_siam}
\begin{figure*}[t]
    \centering
    \safeincludegraphics[width=1.0\textwidth]{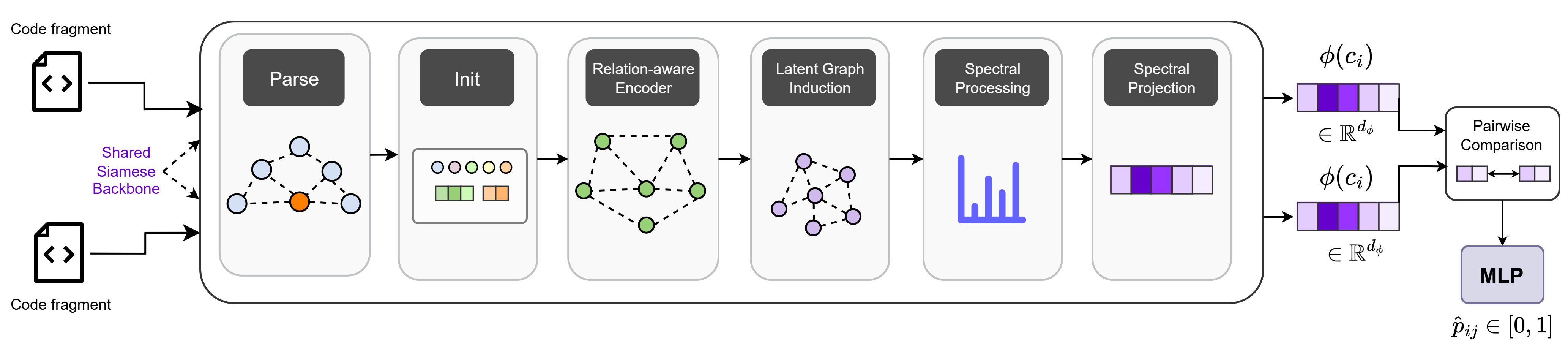}
    \caption{\method{} architecture.}
    \label{fig:spectra_overview}
\end{figure*}

\method{} uses a Siamese architecture with shared parameters for the two code fragments \citep{bromley1993signature}. Each branch constructs a normalized attributed program graph and encodes its syntactic and data-dependence relations. It then maps the variable-size graph to a fixed set of latent nodes to learn their weighted adjacency and derives a multi-scale spectral representation. Spectral features are then compared by a pairwise classifier to output a prediction in $[0,1]$. \autoref{fig:spectra_overview} summarizes the pipeline.

\subsubsection{Stage 1: Initial Representation and Language Normalization}
\label{sec:input_representation}
Each code fragment is first converted into an attributed graph whose nodes have numerical representations and whose edges encode structural relations. Combining syntactic structure with semantic or data-flow relations is common in learned program representations \citep{allamanis2018learning,yamaguchi2014cpg}. We use AST to provide the node backbone and augment it with DDG edges whose endpoints can both be mapped to AST nodes. The edges are made bidirectional before batching. 

\paragraph{Canonical node types}
Parser node names vary across languages and extraction tools, so raw types are mapped to a shared vocabulary $\mathcal{K}_{\mathrm{can}}$: 
\begin{quote}\scriptsize Control\_If, Control\_Loop, Control\_Switch, Control\_Return, Control\_Break, Control\_Generic, Var\_Decl, Func\_Decl, Class\_Decl, Param\_Decl, Assign\_Op, Binary\_Op, Unary\_Op, Call\_Expr, Access\_Expr, Literal\_Num, Literal\_Str, Literal\_Bool, Literal\_Null, Literal\_Generic, Structural\_Block, Identifier\_Context, Type\_Meta, and Canonical\_Unknown. 
\end{quote} 
The mapping covers \texttt{Joern}, Python's AST parser, and the language-specific parser fallbacks used in preprocessing; unmatched types are assigned to \texttt{Canonical\_Unknown}.

\paragraph{Hashed lexical features}
The canonical type alone does not preserve all information carried by the original node label. We therefore keep a small amount of lexical information for each node. Non-numeric labels are split at punctuation and camel-case boundaries~\citep{hill2014empirical}, and literals are normalized. The resulting sub-tokens, together with their character $n_{\mathrm{char}}$-grams, are mapped by a deterministic hash function to $N_{\mathrm{hash}}$ indices \citep{weinberger2009feature}. This keeps the lexical vocabulary at a fixed size regardless of the source language or dataset. We keep at most $r_{\max}$ lexical features per node and ignore labels that contain only line numbers. For node $v$, let $\kappa(t_v)$ be its canonical type and $\ell_{v1},\ldots,\ell_{vr}$ its lexical features. We combine the embedding of the canonical type with the average embedding of these lexical features:
\begin{equation}
h_v^{(0)}
=
\operatorname{LN}\!\left(
E_{\mathrm{can}}(\kappa(t_v))
+
\gamma_{\mathrm{lex}}
\frac{1}{r}
\sum_{q=1}^{r}
E_{\mathrm{lex}}(\ell_{vq})
\right),
\
\gamma_{\mathrm{lex}}=\sigma(g_{\mathrm{lex}}).
\label{eq:node_initialization}
\end{equation}
where $E_{\mathrm{can}}$ and $E_{\mathrm{lex}}$ are learned embedding tables. The learned scalar $\gamma_{\mathrm{lex}}\in(0,1)$ controls how strongly the lexical information contributes to the initial node representation. Layer normalization and dropout are applied to the lexical features during training to improve numerical stability and provide regularization \citep{ba2016layer,srivastava2014dropout}. Graphs are restricted to the first $n_{\max}$ serialized AST nodes; padding is excluded from message passing and reconstruction.

\subsubsection{Stage 2: Relation-Aware Structural Encoding}
\label{sec:relation_encoder}
AST and DDG edges represent different relations and should therefore use separate transformations~\citep{schlichtkrull2018modeling}. Let $H^{(l)}\in\mathbb{R}^{n\times d}$ contain node states at layer $l$ and $A^{(r)}$ denote the adjacency for relation $r\in\mathcal{R}$. Each layer computes
\begin{equation}
\begin{aligned}
U^{(l)}
&=
H^{(l)}W_0^{(l)}
+
\sum_{r\in\mathcal{R}}
\widetilde{A}^{(r)}H^{(l)}W_r^{(l)},\\
H^{(l+1)}
&=
\operatorname{LN}\!\left[
H^{(l)}
+
\operatorname{Dropout}
\bigl(\operatorname{GELU}(U^{(l)})\bigr)
\right],
\end{aligned}
\label{eq:relation_layer}
\end{equation}
where $\widetilde{A}^{(r)}$ is row-normalized, with a minimum degree denominator of one. $W_0^{(l)}$ transforms the current state and $W_r^{(l)}$ is relation specific. The residual, normalization, GELU, and dropout components follow standard formulations \citep{he2016deep,ba2016layer,hendrycks2016gaussian,srivastava2014dropout}. The encoder uses $L_{\mathrm{enc}}$ layers of dimension $d$.

\subsubsection{Stage 3: Latent Graph Induction}
\label{sec:latent_induction}
We reduce the number of node states to $m$ latent nodes using iterative soft assignment, following the general idea of differentiable graph pooling
\citep{ying2018hierarchical}. For normalized input key $k_v$ and latent query $q_j$,
\begin{equation}
P_{vj}
=
\frac{\exp(q_j^{\top}k_v/\sqrt{d})}
{\sum_{u=1}^{m}\exp(q_u^{\top}k_v/\sqrt{d})}.
\label{eq:slot_assignment}
\end{equation}
Latent updates normalize contributions over valid input nodes and use a GRU followed by a residual MLP~\citep{cho2014gru,he2016deep}. After $I_{\mathrm{assign}}$ iterations, the latent states form $Z\in\mathbb{R}^{m\times d}$. To limit information loss during pooling, $\widehat{H}=PZ$ reconstructs the input states. With $M_v$ indicating valid nodes,
\begin{equation}
\mathcal{L}_{\mathrm{rec}}
=
\frac{1}{\sum_v M_v}
\sum_v
M_v
\frac{1}{d}
\left\|
h_v^{(0)}-\widehat{h}_v
\right\|_2^2.
\label{eq:reconstruction_loss}
\end{equation}
The observed structure is projected into latent space as
\begin{equation}
B=P^{\top}A_{\mathrm{in}}P,
\qquad
\widetilde{B}
=
\frac{B}{\max_{i,j}B_{ij}+\epsilon},
\label{eq:structure_projection}
\end{equation}
where $A_{\mathrm{in}}$ combines the enabled input relations and $\widetilde{B}$ serves as a prior. Task-dependent affinities are obtained from $H_a$ scaled dot-product attention heads~\citep{vaswani2017attention}:
\begin{equation}
S
=
\operatorname{Sym}\!\left(
\frac{1}{H_a}
\sum_{h=1}^{H_a}
\frac{Q_hK_h^{\top}}{\sqrt{d_h}}
\right)
+
\eta\widetilde{B},
\label{eq:latent_affinity}
\end{equation}
where $\eta$ learns the contribution of the projected AST/DDG structure. Thus, the latent topology combines observed structure with task-dependent affinities, following the general setting of graph structure learning~\citep{jin2020graph}.

\begin{figure}[!htbp]
    \centering
    \safeincludegraphics[width=0.9\columnwidth]{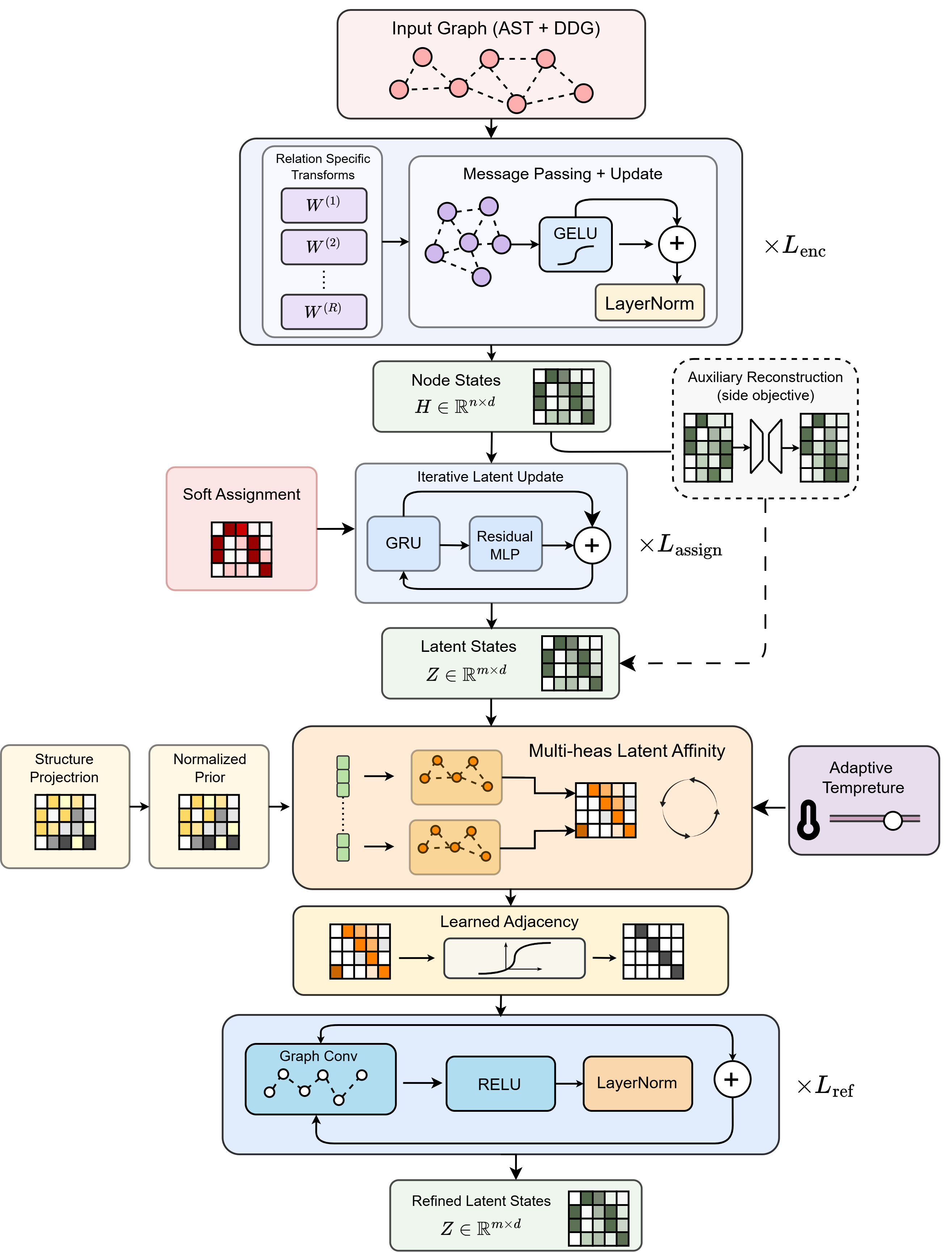}
    \caption{
    Latent graph induction in \method{}.
    }
    \label{fig:latent_graph_induction}
\end{figure}

\paragraph{Adaptive edge temperature}
We convert $S$ to continuous edge weights using a temperature conditioned on graph size and density. For $n(c)$ valid nodes, define node occupancy $u(c)=n(c)/n_{\max}$ and
$\rho_{\mathrm{in}}(c)=\operatorname{clip}_{[0,1]}(\sum_{r,i,j}A_{ij}^{(r)}/n(c)^2)$.
Their mixture is
$\chi(c)=\beta_u u(c)+(1-\beta_u)\rho_{\mathrm{in}}(c)$, giving
\begin{equation}
T(c)
=
T_{\min}
+
(T_{\max}-T_{\min})
\sigma\!\left(
b_T+\operatorname{softplus}(w_T)\chi(c)
\right).
\label{eq:adaptive_temperature}
\end{equation}
After clipping $S$ to $[-s_{\max},s_{\max}]$, the latent adjacency is
\begin{equation}
A_{\Theta}(c)
=
\operatorname{Sym}\!\left[
\sigma\!\left(\frac{S}{T(c)}\right)
\odot
(\mathbf{1}\mathbf{1}^{\top}-I_m)
\right].
\label{eq:latent_adjacency}
\end{equation}
$A_{\Theta}(c)$ is symmetric, continuous, and zero-diagonal; no hard edge threshold is applied. $L_{\mathrm{ref}}$ additional graph layers refine $Z$ using this learned adjacency. \autoref{fig:latent_graph_induction} summarizes the complete latent-graph induction process.

\subsubsection{Stage 4: Multi-Scale Spectral Representation}
\label{sec:multiscale_spectrum}
The learned graph and its latent-node attributes are summarized jointly through an attributed graph-spectral representation. We first construct the normalized Laplacian
\begin{equation}
L_{\Theta}
=
I_m-D_{\Theta}^{-1/2}A_{\Theta}D_{\Theta}^{-1/2},
\qquad
D_{\Theta}
=
\operatorname{diag}(A_{\Theta}\mathbf{1}),
\label{eq:normalized_laplacian}
\end{equation}
whose eigenvalues lie in $[0,2]$ for an undirected graph with non-negative edge weights \citep{chung1997spectral}. Numerical stability is maintained by exact symmetrization of the Laplacian, clamping the spectrum to its theoretical range [0, 2], and forcing single precision for the eigensolver under mixed-precision training.

\paragraph{Eigenvalue density}
For normalized Laplacian eigenvalues
$\lambda_1,\ldots,\lambda_m$, a Gaussian-smoothed profile
\citep{banerjee2008spectrum,lewitus2016characterizing} is evaluated at $N_\xi$ centers $\xi_j\in[0,2]$:
\begin{equation}
d_j
=
\frac{1}{m}
\sum_{i=1}^{m}
\exp\!\left[
-\frac{1}{2}
\left(
\frac{\lambda_i-\xi_j}{\sigma_{\mathrm{dens}}}
\right)^2
\right],
\quad
j=1,\ldots,N_{\xi},
\label{eq:eigenvalue_density}
\end{equation}
giving $d(c)\in\mathbb{R}^{N_{\xi}}$.

\paragraph{Heat traces}
Heat diffusion captures structure across scales
\citep{sun2009concise}. At logarithmically spaced $t_q\in[t_{\min},t_{\max}]$,
\begin{equation}
h_q
=
\frac{1}{m}
\sum_{i=1}^{m}
\exp(-t_q\lambda_i),
\qquad
q=1,\ldots,N_t,
\label{eq:heat_trace}
\end{equation}
yielding $h(c)\in\mathbb{R}^{N_t}$.

\paragraph{Graph-signal Chebyshev energies}
To preserve information about node attributes within the spectral representation, the refined latent states $Z$ are projected to $N_{\mathrm{sig}}$ graph-signal channels:
\begin{equation}
\widetilde{X}
=
ZW_s
\in\mathbb{R}^{m\times N_{\mathrm{sig}}}.
\label{eq:raw_graph_signal}
\end{equation}
The signals are RMS-normalized across latent nodes:
\begin{equation}
X_{vr}
=
\frac{\widetilde{X}_{vr}}
{
\sqrt{
\frac{1}{m}
\sum_{u=1}^{m}
\widetilde{X}_{ur}^{\,2}
+\epsilon
}
},
\qquad
r=1,\ldots,N_{\mathrm{sig}}.
\label{eq:graph_signal}
\end{equation}
This normalization removes signal-scale effects before spectral filtering~\citep{stankovic2022introduction}. Since the normalized-Laplacian spectrum lies in $[0,2]$, we use $\widetilde{L}=L_{\Theta}-I_m$ and the Chebyshev recurrence
\begin{equation}
T_0(X)=X,\quad
T_1(X)=\widetilde{L}X,\quad
T_k(X)=2\widetilde{L}T_{k-1}(X)-T_{k-2}(X).
\end{equation}
Chebyshev approximations allow spectral filtering without explicit eigenvector multiplication \citep{defferrard2016chebnet}. For $N_{\mathrm{band}}$ Gaussian filters,
\begin{equation}
Y_b
=
\sum_{k=0}^{K_{\mathrm{cheb}}}
a_{bk}T_k(X),
\qquad
b=1,\ldots,N_{\mathrm{band}},
\label{eq:chebyshev_filter}
\end{equation}
where the filter centers lie in $[\omega_{\min},\omega_{\max}]$
with bandwidth $\sigma_{\mathrm{filt}}$ and fixed coefficients $a_{bk}$. Gradients propagate through the eigendecomposition, so the eigenvalue-density and heat-trace profiles supervise the learned adjacency alongside the differentiable Chebyshev path. Band importance is fixed at $\alpha_{b}=1/N_{\mathrm{band}}$, and the per-band energies are
\begin{equation}
e_{br}
=
\log\big(1+\frac{1}{m}\sum_{i=1}^{m}Y_b(i,r)^2\big).
\label{eq:spectral_energy}
\end{equation}
Thus $e(c)\in\mathbb{R}^{N_{\mathrm{band}}N_{\mathrm{sig}}}$, and the complete descriptor is
\begin{equation}
s_{\Theta}(c)
=
d(c)\concat h(c)\concat e(c)
\in\mathbb{R}^{d_s},
\
d_s
=
N_{\xi}+N_t+N_{\mathrm{band}}N_{\mathrm{sig}}.
\label{eq:spectral_descriptor}
\end{equation}

\subsubsection{Spectral Embedding and Pairwise Prediction}
\label{sec:head}
We compute $\phi_{\Theta}(c)=\operatorname{Norm}_2\!\left(g_s(s_{\Theta}(c))\right)\in\mathbb{R}^{d_{\phi}},$ where $g_s$ is a learned projection. For pair $(c_i,c_j)$, with $\phi_i=\phi_{\Theta}(c_i)$ and $\phi_j=\phi_{\Theta}(c_j)$, we form
\begin{equation}
q_{ij}
=
\phi_i
\concat\phi_j
\concat|\phi_i-\phi_j|
\concat(\phi_i\odot\phi_j)
\concat\cos(\phi_i,\phi_j),
\label{eq:pair_vector}
\end{equation}
where $q_{ij}\in\mathbb{R}^{4d_{\phi}+1}$. An $L_{\mathrm{pair}}$-layer MLP maps $q_{ij}$ to a scalar logit, and a sigmoid yields $\widehat{p}_{ij}$.

\subsubsection{Training Objective}
\label{sec:training_objective}
Training combines classification, spectral contrast, AUC ranking, hard-negative separation, node and topology reconstruction, spectral variance preservation, and graph regularization.

\paragraph{Classification loss}
With $N_+$ and $N_-$ denoting clone and non-clone training pairs, define $w_+=\min(w_{\max},N_-/\max(1,N_+))$. For mini-batch $\mathcal{B}$,
\begin{equation}
\mathcal{L}_{\mathrm{cls}}
=
-\frac{1}{|\mathcal{B}|}
\sum_{(i,j)\in\mathcal{B}}
\left[
w_+y_{ij}\log\widehat{p}_{ij}
+
(1-y_{ij})\log(1-\widehat{p}_{ij})
\right].
\label{eq:classification_loss}
\end{equation}

\paragraph{Spectral contrastive loss}
For $c^s_{ij}=\cos(s_{\Theta}(c_i),s_{\Theta}(c_j))$, contrastive supervision \citep{hadsell2006contrastive} is
\begin{equation}
\mathcal{L}_{\mathrm{spec}}
=
\frac{1}{|\mathcal{B}|}
\sum_{(i,j)\in\mathcal{B}}
\left[
y_{ij}(1-c^s_{ij})^2
+
(1-y_{ij})
\relu{c^s_{ij}-\mu_{\mathrm{spec}}}^{2}
\right].
\label{eq:spectral_contrastive}
\end{equation}

\paragraph{AUC-ranking loss}
Let $r_{ij}$ denote the classifier logit before the sigmoid, and let
$\mathcal{B}_{+}$ and $\mathcal{B}_{-}$ contain the positive and negative
pairs in the mini-batch, respectively. The pairwise logistic loss ranks
positive pairs above negatives:
\begin{equation}
\mathcal{L}_{\mathrm{auc}}
=
\frac{1}{|\mathcal{B}_{+}|\,|\mathcal{B}_{-}|}
\sum_{(i,j)\in\mathcal{B}_{+}}
\sum_{(u,v)\in\mathcal{B}_{-}}
\log\!\left(1+\exp\!\left[-(r_{ij}-r_{uv})\right]\right).
\label{eq:auc_ranking_loss}
\end{equation}
This term is set to zero when either class is absent from the mini-batch.

\paragraph{Hard-negative loss} To emphasize confusable non-clones
\citep{wang2022disco}, let
$c^{\phi}_{ij}=\cos(\phi_i,\phi_j)$ and let $\mathcal{K}$ index the
$\max(1,\lfloor\kappa_{\mathrm{hard}}N_-^{\mathcal{B}}\rfloor)$ largest negative-pair values of
$\relu{c^{\phi}_{ij}-\mu_{\mathrm{hard}}}^{2}$. Then
\begin{equation}
\mathcal{L}_{\mathrm{hard}}
=
\frac{1}{|\mathcal{K}|}
\sum_{(i,j)\in\mathcal{K}}
\relu{
c^{\phi}_{ij}-\mu_{\mathrm{hard}}
}^{2}.
\label{eq:hard_negative}
\end{equation}

\paragraph{Topology-reconstruction loss}
Let $R=PA_{\Theta}P^{\top}$ be the learned adjacency projected back to the
input-node space and let $T=\max(A_{\mathrm{in}},A_{\mathrm{in}}^{\top})$ be
the symmetrized AST/DDG adjacency. Over the valid off-diagonal entries
$\Omega$, class-balanced binary cross-entropy gives
\begin{equation}
\mathcal{L}_{\mathrm{topo}}
=
-\frac{1}{|\Omega|}
\sum_{(u,v)\in\Omega}
\left[
w_{\mathrm{topo}}T_{uv}\log R_{uv}
+
(1-T_{uv})\log(1-R_{uv})
\right],
\label{eq:topology_reconstruction_loss}
\end{equation}
where $w_{\mathrm{topo}}$ is the clipped ratio of non-edges to edges in
$\Omega$. This term encourages $PA_{\Theta}P^{\top}$ to reproduce the input
AST/DDG edges.

\paragraph{Spectral-variance loss}
Let $\sigma_k$ be the across-batch standard deviation of coordinate $k$ of
the $\ell_2$-normalized spectral descriptors from both sides of all pairs.
The variance-preserving hinge loss is
\begin{equation}
\mathcal{L}_{\mathrm{var}}
=
\relu{
0.03-\frac{1}{d_s}\sum_{k=1}^{d_s}\sigma_k
}.
\label{eq:spectral_variance_loss}
\end{equation}
It keeps the mean descriptor standard deviation above $0.03$ and prevents
spectral collapse.

\paragraph{Graph regularization}
For a batch of latent adjacencies, let
\begin{equation}
\bar{a}(A)
=
\frac{
\sum_{b=1}^{|\mathcal{B}|}\sum_{i\neq j}A_{b,ij}
}{
|\mathcal{B}|m(m-1)
},
\qquad
\mathcal{L}_{\mathrm{den}}
=
\bigl(\bar{a}(A)-\bar{a}_0\bigr)^2.
\end{equation}
We penalize low algebraic connectivity based on the second-smallest Laplacian eigenvalue, i.e., algebraic connectivity \citep{fiedler1973algebraic}:
\begin{align}
&\mathcal{L}_{\mathrm{conn}}
=
\frac{1}{|\mathcal{B}|}
\sum_b
\relu{\mu_{\mathrm{conn}}-\lambda_2^{(b)}}^2,
\\
&\mathcal{L}_{\mathrm{graph}}
=
\alpha_{\mathrm{den}}\mathcal{L}_{\mathrm{den}}
+
\alpha_{\mathrm{conn}}\mathcal{L}_{\mathrm{conn}}.
\label{eq:graph_regularizer}
\end{align}
The complete objective is
\begin{equation}
\begin{aligned}
\mathcal{L}_{\mathrm{total}}
={}&\mathcal{L}_{\mathrm{cls}}
+\lambda_{\mathrm{spec}}\mathcal{L}_{\mathrm{spec}}
+\lambda_{\mathrm{hard}}\mathcal{L}_{\mathrm{hard}}\\
&+\lambda_{\mathrm{rec}}\mathcal{L}_{\mathrm{rec}}
+\lambda_{\mathrm{graph}}\mathcal{L}_{\mathrm{graph}}\\
&+\lambda_{\mathrm{auc}}\mathcal{L}_{\mathrm{auc}}
+\lambda_{\mathrm{topo}}\mathcal{L}_{\mathrm{topo}}
+\lambda_{\mathrm{var}}\mathcal{L}_{\mathrm{var}}.
\end{aligned}
\label{eq:complete_loss}
\end{equation}

\subsubsection{Optimization}
\label{sec:optimization}

All learnable components are optimized with AdamW
\citep{loshchilov2019adamw}. Neural layers use mixed-precision GPU
computation \citep{micikevicius2018mixed}, while latent-adjacency
construction and spectral operations use single precision for
numerical stability. Numerical hyperparameters are reported in \autoref{tab:spectra_settings} of \autoref{sec:settings}. Checkpoint and decision-threshold selection use only the validation split (see \autoref{sec:protocol}).

\section{Experimental Design}
\label{sec:experimental_design}

\subsection{Research Questions}
\label{sec:research_questions}
The objective of this study is to determine whether learning a latent graph can improve spectral code representation for effective and generalizable cross-language clone detection. To achieve this objective, we formulate three main research questions as follows:

\begin{itemize}[leftmargin=*]
\item \textbf{RQ1}
\emph{To what extent does learning a latent graph improve the discriminative power of spectral code representations compared with fixed program graphs?}

\item \textbf{RQ2}
\emph{How effective is \method{} compared with non-graph, graph-based, hybrid graph--code, and pre-trained baselines under conventional evaluation?}

\item \textbf{RQ3}
\emph{How well does \method{} generalize to unseen language combinations when target-language supervision is unavailable?}

\end{itemize}

\subsection{Dataset}
\label{sec:datasets}
We use three semantic clone datasets: BigCloneBench~\citep{svajlenko2014bigclonebench},
filtered following \citet{wang2020faast}, AtCoder~\citep{perez2019crosslang},
and a benchmark derived from Project CodeNet~\citep{puri2021project}. BigCloneBench contains Java--Java pairs, while AtCoder contains Java--Python pairs. CodeNet covers four within-language configurations (Python--Python, Java--Java, C++--C++, and C\#--C\#) and all six cross-language configurations (Python--Java, Python--C++, Python--C\#, Java--C++, Java--C\#, and C++--C\#). The final CodeNet benchmark is balanced both by language configuration and clone label. Each of the ten configurations contains 10K pairs, with 5K clones and 5K non-clones, for 100K pairs in total. Non-clone pairs for AtCoder and CodeNet are formed from accepted solutions to different programming problems, following~\citep{perez2019crosslang}. \autoref{tab:datasets} summarizes the resulting datasets and data splits.

\begin{table}[!htbp]
\centering
\caption{Datasets used in the evaluation. CodeNet contains ten balanced language configurations, each with 10K pairs and 5K clones.}
\label{tab:datasets}

\scriptsize
\renewcommand{\arraystretch}{1.08}
\setlength{\tabcolsep}{2.5pt}

\begin{tabular}{|l|r|r|r|r|r|}
\hline
\textbf{Dataset} & \textbf{Total} & \textbf{Clones} &
\textbf{Train} & \textbf{Val.} & \textbf{Test} \\
\hline

\makecell[l]{BigCloneBench\\\textit{Java--Java}}
& 1,731,860
& 561,521
& 901,028
& 415,416
& 415,416
\\
\hline

\makecell[l]{AtCoder\\\textit{Java--Python}}
& 777,494
& 388,747
& 544,054
& 116,704
& 116,736
\\
\hline

\makecell[l]{CodeNet\\\textit{10 configurations}}
& 100,000
& 50,000
& 70,000
& 15,000
& 15,000
\\
\hline

\end{tabular}
\end{table}

\paragraph{CodeNet's Clone Extraction Process} We construct CodeNet pairs from problem IDs and submission verdicts. Two distinct accepted submissions to the same problem form a clone, while accepted submissions to different problems form a non-clone. We first build a diversity-controlled pool of approximately one million pairs. For each language configuration, candidate problems are sampled across eight groups based on the number of pairs they can provide, and submissions are selected to vary in user, submission time, code size, and compiler variant. For same-language pairs, near-duplicates with Jaccard similarity of at least 0.90 are removed using normalized token shingles and MinHash/LSH \citep{broder1997resemblance,broder1998minwise}; less-similar and less-used submissions are preferred, and each submission is limited to at most ten selected pairs. From this pool, we sample the final benchmark by language configuration and clone label, yielding 5K clone and 5K non-clone pairs for each of the ten configurations. The resulting 100K pairs are then split into 70K training, 15K validation, and 15K test pairs while preserving this stratification.

\subsection{Baselines}
\label{sec:baselines}
We compare \method{} with six baseline families: fixed spectral representations, GNNs over conventional program graphs, non-graph code models, other graph-based models, a hybrid graph--code model, and pre-trained code models. For the fixed spectral representations, we consider four downstream settings: no training (\textit{No Train}), Random Forest (RF) \citep{breiman2001random}, Logistic Regression (LR) \citep{hastie2009elements}, and a Siamese Neural Network (SNN) \citep{bromley1993signature}. Combined with AST, CFG, DDG, and CPG, this gives 16 fixed-spectral configurations. The GNN~\citep{gilmer2017neural} baselines use the same four graph views. We further evaluate Deckard~\citep{jiang2007deckard}, RtvNN~\citep{white2016deep}, CDLH~\citep{wei2017cdlh}, ASTNN~\citep{zhang2019astnn}, FA-AST+GGNN~\citep{wang2020faast}, FA-AST+GMN~\citep{wang2020faast}, and DeepSim~\citep{zhao2018deepsim}. Finally, CodeBERT~\citep{feng2020codebert} and GraphCodeBERT~\citep{guo2021graphcodebert} are evaluated using \emph{No Train}, RF, SNN~\citep{bromley1993signature}, and Principal Component Analysis (PCA)~\citep{jolliffe2005principal} followed by RF, and PCA followed by SNN (PCA+SNN). Their main hyperparameters/configurations are summarized in \autoref{tab:baselines}.

\begin{table}[!htbp]
\centering
\caption{Baseline configurations. For neural models, the training column reports batch size / learning rate $\alpha$ / maximum epochs.}
\label{tab:baselines}

\tiny
\renewcommand{\arraystretch}{1.12}
\setlength{\tabcolsep}{1.6pt}

\renewcommand{\tabularxcolumn}[1]{m{#1}}

\begin{tabularx}{\columnwidth}{@{}
m{0.19\columnwidth}
m{0.34\columnwidth}
X
r@{}}
\toprule
Method & Input / architecture & Training & Params. \\
\midrule

\rowcolor{black!7}
\multicolumn{4}{@{}l}{\textit{Fixed spectral representations (AST, CFG, DDG, CPG)}} \\[1pt]

No Train
&
136-D spectrum
&
No training
&
0
\\
\midrule

RF
&
274-D pair features
&
\shortstack[l]{200 trees; depth 16; leaf 5\\
balanced-subsample; seed 42}
&
--
\\
\midrule

LR
&
274-D pair features
&
\shortstack[l]{train-only scaling; balanced\\
max.\ 1,000 iter.; seed 42}
&
--
\\
\midrule

SNN
&
$136\!\to\!256\!\to\!256\!\to\!128$
&
\shortstack[l]{$8192/10^{-3}/4$; dropout .10\\
WD $10^{-4}$; clip 5.0}
&
202k
\\
\midrule

\rowcolor{black!7}
\multicolumn{4}{@{}l}{\textit{GNNs over conventional graphs (AST, CFG, DDG, CPG)}} \\[1pt]

GCN
&
\shortstack[l]{5-D nodes; GCN $256/128$; mean pool\\
$128+128+128=384$-D representation}
&
\shortstack[l]{$16/(5{\times}10^{-4})/4$; dropout .05\\
WD $10^{-4}$; patience 2; clip 2.0}
&
989k
\\
\midrule

\rowcolor{black!7}
\multicolumn{4}{@{}l}{\textit{Token, tree, graph, and hybrid baselines}} \\[1pt]

Deckard
&
\shortstack[l]{AST-category count vectors per subtree\\
top-32 subtrees by size; cosine}
&
Validation threshold
&
0
\\
\midrule

RtvNN
&
\shortstack[l]{256 AST nodes; recursive tree encoder\\
emb.\ 128; hidden 192; 192-D}
&
\shortstack[l]{$32/(5{\times}10^{-4})/4$; dropout .15\\
WD $10^{-4}$; patience 2; clip 1.0}
&
1.48--4.46M
\\
\midrule

CDLH
&
\shortstack[l]{256 AST nodes; binary Tree-LSTM\\
emb.\ 128; hidden 192; 32-bit hash}
&
\shortstack[l]{$16/10^{-3}/4$; WD 0\\
patience 2; clip 1.0}
&
2.10--6.96M
\\
\midrule

ASTNN
&
\shortstack[l]{256 nodes; 512 edges; 64 statement roots\\
emb.\ 128; recursive subtree\\ + BiGRU 100; 200-D}
&
\shortstack[l]{$32/10^{-3}/4$; dropout .20\\
WD 0; patience 2; clip 1.0}
&
1.15--2.16M
\\
\midrule

FA-AST+GGNN
&
\shortstack[l]{256 nodes; 6 relations\\ (AST/CFG/DDG/CPG)\\
hidden 192; 5 steps; gated readout; 192-D}
&
\shortstack[l]{$8/(5{\times}10^{-4})/4$; dropout .10\\
WD $10^{-4}$; patience 2; clip 1.0}
&
556k
\\
\midrule

FA-AST+GMN
&
\shortstack[l]{as FA-AST+GGNN, plus cross-graph\\
attention at every propagation step}
&
\shortstack[l]{$4/(5{\times}10^{-4})/4$; dropout .10\\
WD $10^{-4}$; patience 2; clip 1.0}
&
--
\\
\midrule

DeepSim
&
\shortstack[l]{CFG/DDG feature matrix\\
192-D representation}
&
\shortstack[l]{$32/(5{\times}10^{-4})/4$; WD $10^{-3}$\\
patience 2; clip 1.0}
&
--
\\

\bottomrule
\end{tabularx}
\end{table}

For the fixed spectral baselines, each AST, CFG, DDG, or CPG spectrum is sorted, zero-padded or truncated to 64 eigenvalues, and augmented with eight statistics: normalized spectral length $\min(|\lambda|/2000,10)$, mean, standard deviation, minimum, maximum, and the 25th, 50th, and 75th percentiles. This gives the 72-dimensional representation $x_b(c)=
\operatorname{Stat}\!\left(\lambda(G_b(c))\right)
\concat
\lambda_{1:64}\!\left(G_b(c)\right)$.
\emph{No Train} scores a pair by
$s_b(c_i,c_j)=
1/(1+\lVert x_b(c_i)-x_b(c_j)\rVert_2)$,
whereas RF and LR use the 146-dimensional vector
\begin{equation}
r_{ij}
=
|x_i-x_j|
\concat(x_i\odot x_j)
\concat\cos(x_i,x_j)
\concat\lVert x_i-x_j\rVert_2.
\label{eq:fixed_pair_features}
\end{equation}

The GNN node features consist of a constant value, log
in-degree, log out-degree, log total degree, and normalized
serialized position. Deckard hashes normalized tokens, token bigrams, token- and line-count buckets, nesting depth, returns, branches, and loops. ASTNN applies two bottom-up tree-convolution layers before its statement-level BiGRU. FA-AST+GGNN and FA-AST+GMN use CPG as the available flow-augmented-AST proxy. DeepSim combines its CFG branch with source-code semantic/token features. All neural baselines implemented in our controlled repository use binary cross entropy (BCE) with logits, AdamW, validation-F1 early stopping, and mixed precision GPU training.

\subsection{Experimental Protocol}
\label{sec:protocol}
Each method uses the train, validation, and test partitions given in \autoref{tab:datasets} of \autoref{sec:datasets}. Models are fitted on the training pairs, while all hyperparameter choices, checkpoint selection, and decision-threshold tuning are performed using only the training and validation splits. For the approximately balanced AtCoder and CodeNet datasets, validation accuracy is used as the model-selection criterion. For the imbalanced BigCloneBench dataset, we instead maximize binary F1 on the validation set. The test split is accessed only once for final performance reporting after model selection is complete.

For RQ1, we also evaluate the spectral representation learned by \method{} independently of its original MLP head in \autoref{sec:head}, using \emph{No Train} RF, LR, and SNN as downstream predictors. For RQ2, we additionally evaluate three nested variants of \method{} to separate the contribution of structural topology, canonical node labels, and lexical information. \method{} \emph{(Topo.)} disables both the canonical node-type and hashed lexical features introduced in \autoref{sec:input_representation}. \method{} \emph{(Label)} adds the canonical node-type representation but keeps the hashed lexical features disabled. \method{} \emph{(Lex.)} is the complete model and uses both the canonical type and hashed lexical components of the node representation in \autoref{eq:node_initialization}. The three variants are otherwise identical.

For the CodeNet experiments in RQ3 (\autoref{sec:rq3}), J, P, C, and S denote Java, Python, C++, and C\#, respectively, and $Y$ denotes the target language. We consider three settings. In the \emph{within-language} setting, training uses $Y$--$Y$ pairs. In the \emph{endpoint-only} setting, a second language $X$ is selected; training uses $X$--$X$ and $Y$--$Y$ pairs, and evaluation is performed on the unseen $X$--$Y$ pair. In the \emph{bridge-assisted} setting, $X_1,X_2,X_3$ denote distinct non-target languages forming a path toward $Y$. Training includes $X_1$--$X_1$ and the consecutive cross-language pairs indicated by the path ($X_1$--$X_2$,$X_2$--$X_3$, $X_3$--$Y$), with optional same-language pairs for intermediate languages ($X_2$--$X_2$,$X_3$--$X_3$). Within-language and bridge-assisted settings are evaluated on $Y$--$Y$ test pairs.

\subsection{Experimental Settings}
\label{sec:settings}
Learning experiments run in multiple Kaggle notebooks with a T4-class CUDA accelerator. Graph construction (AST, CFG, DDG, CPG) and dataset preparation are performed beforehand, using the repository's local Python/\texttt{Joern} pipeline, after which the cleaned data bundles are attached to Kaggle.

We use a fixed canonical configuration for \method{} throughout the main experiments. This configuration specifies the graph input and latent construction, spectral representation, pair encoder, regularization terms, and optimization settings. The complete configuration is reported in \autoref{tab:spectra_settings}.  

Our evaluation includes 893 neural training runs across different datasets, model variants, baselines, and cross-language settings. This gives us broad experimental coverage, but it also makes long training runs expensive. We therefore cap \method{} and the neural baselines at 4 epochs in the main evaluation, so that all methods are compared under a similar training budget. \autoref{sec:num_epochs} shows this cap penalizes \method{}; its CodeNet accuracy rises from .68 at four epochs to .77 at twenty, so the main tables in \autoref{sec:exp_results} report a conservative lower bound on our method.

\begin{table}[!htbp]
\centering
\caption{Canonical \method{} architecture and training settings.}
\label{tab:spectra_settings}

\scriptsize
\renewcommand{\arraystretch}{1.1}
\setlength{\tabcolsep}{2pt}

\begin{tabularx}{\columnwidth}{
@{}
>{\raggedright\arraybackslash}X
>{\raggedleft\arraybackslash}p{0.37\columnwidth}
@{}}
\toprule
{Setting (symbol)} & {Value} \\
\midrule
\multicolumn{2}{@{}l}{\textit{{Architecture and representation}}} \\
{Input graph ($n_{\max}$; relations)}
& {256 nodes; AST+DDG} \\
{Node typing/hash
($|\mathcal K_{\mathrm{can}}|,n_{\mathrm{char}},N_{\mathrm{hash}},r_{\max}$)}
& {24; 3; 4,096; 4} \\
{Lexical features
($\gamma_{\mathrm{lex}}^{(0)},p_{\mathrm{lex}}$)}
& {0.20; 0.30} \\
{Structural encoder
($d,L_{\mathrm{enc}},p_{\mathrm{enc}}$)}
& {256; 2; 0.10} \\
{Latent assignment
($m,I_{\mathrm{assign}}$)}
& {32; 3} \\
{Latent affinity/refinement
($H_a,\eta^{(0)},L_{\mathrm{ref}}$)}
& {4; 1.0; 2} \\
{Adaptive temperature
($\beta_u,T_{\min},T_{\max}$)}
& {0.75; 0.20; 1.20} \\
{Temperature init./clipping
($b_T^{(0)},w_T^{(0)},s_{\max}$)}
& {$-0.55$; 1.25; 20} \\
{Chebyshev order ($K_{\mathrm{cheb}}$)}
& {12} \\
{Eigenvalue density kernel
($\sigma_{\mathrm{dens}}$)}
& {0.08} \\
{Heat times
($t_{\min},t_{\max}$)}
& {$10^{-2}$; $10^{2}$} \\
{Chebyshev filter centers
($\omega_{\min},\omega_{\max},\sigma_{\mathrm{filt}}$)}
& {0.05; 1.95; 0.18} \\
{Spectral density
($N_{\xi},N_{t},N_{\mathrm{band}},N_{\mathrm{sig}}$)}
& {32; 24; 12; 8} \\
{Spectral descriptor ($d_s$)}
& {152} \\
{Graph-signal floor ($\epsilon$)}
& {$10^{-6}$} \\
{Pair encoder
($d_{\phi},L_{\mathrm{pair}}$)}
& {256; 3} \\
{Trainable parameters ($m=32$)}
& {3,016,855} \\
\midrule
\multicolumn{2}{@{}l}{\textit{{Training and regularization}}} \\
{Class/similarity settings
($w_{\max},\mu_{\mathrm{spec}}$)}
& {10; 0.25} \\
{Hard-negative mining
($\kappa_{\mathrm{hard}},\mu_{\mathrm{hard}}$)}
& {0.25; 0.10} \\
{Graph regularization
($\bar a_0,\mu_{\mathrm{conn}}$)}
& {0.15; 0.03} \\
{Graph-reg.\ weights
($\alpha_{\mathrm{den}},\alpha_{\mathrm{conn}}$)}
& {2; 0.1} \\
{Topology-loss class weight
($w_{\mathrm{topo}}$)}
& {clipped to $[1,20]$} \\
{Loss weights ($\lambda_{\mathrm{spec}},\lambda_{\mathrm{hard}},\lambda_{\mathrm{rec}},\lambda_{\mathrm{graph}},\lambda_{\mathrm{auc}},\lambda_{\mathrm{topo}},\lambda_{\mathrm{var}}$)}
& {0.30; 0.20; 0.05; 0.01; 0.10; 0.05; 0.05} \\
{Optimizer}
& {AdamW} \\
{Learning rate / weight decay
($\eta_{\mathrm{lr}},\lambda_{\mathrm{wd}}$)}
& {$10^{-4}$; $10^{-4}$} \\
{Batch / accumulation
($B,N_{\mathrm{acc}},BN_{\mathrm{acc}}$)}
& {32; 4; 128} \\
{Epochs; seed}
& {4; 42} \\
\bottomrule
\end{tabularx}
\end{table}

\subsection{Evaluation Metrics}
\label{sec:metrics}
We treat clone pairs as the positive class and report precision, recall, F1, and accuracy:
\begin{align}
P &= \frac{TP}{TP+FP},
&
R &= \frac{TP}{TP+FN},
\nonumber\\
F1 &= \frac{2PR}{P+R},
&
\mathrm{Acc.} &= \frac{TP+TN}{TP+TN+FP+FN}.
\label{eq:evaluation_metrics}
\end{align}
Here, $TP$, $TN$, $FP$, and $FN$ denote true positives, true negatives, false positives, and false negatives, respectively.

\section{Experimental Results}
\label{sec:exp_results}

\subsection{RQ1: Discriminative Power of the Learned Spectral Representation}
\label{sec:rq1}
\autoref{tab:rq1} compares the learned latent spectrum of \method{} with the spectra derived from conventional program graphs under the same downstream settings. On BigCloneBench, the naive \emph{Predict All Clone} baseline already obtains .24 F1, and the untrained AST, CFG, DDG, and CPG spectra remain at essentially the same level (.24--.25), indicating little useful separation between clone and non-clone pairs. Training downstream classifiers like SNN improves these representations, but the best fixed-graph configuration reaches only .37 F1. In contrast, the spectrum learned by \method{} reaches .50 F1 with no downstream training at all, and rises to .67 when used with an SNN, despite removing the original prediction head of \method{} in \autoref{sec:head}.

\begin{figure}[!htbp]
    \centering

    \subfloat[\method{} Spectrum + No Train.]{
        \includegraphics[
            width=0.95\columnwidth,
        ]{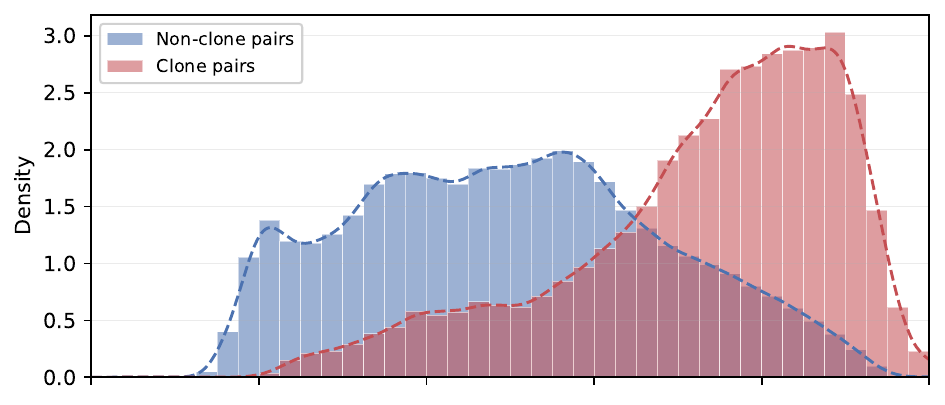}
        \label{fig:rq1_method}
    }
    
    \vspace{2mm}
    
    \subfloat[AST + No Train.]{
        \includegraphics[
            width=0.95\columnwidth,
        ]{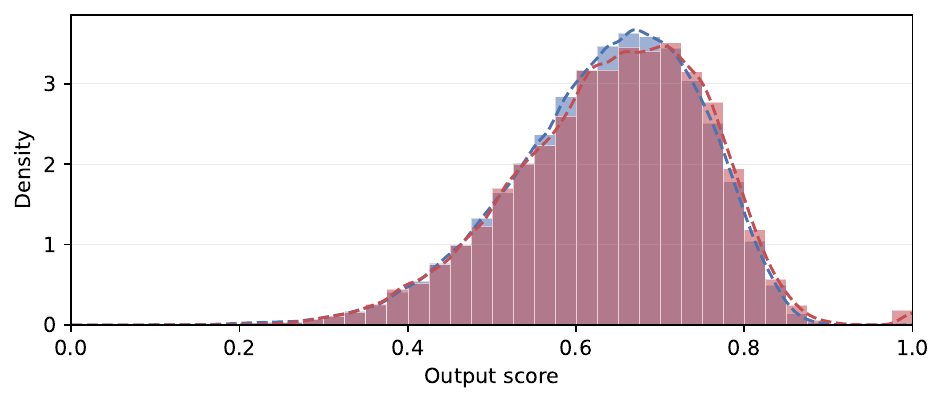}
        \label{fig:rq1_best_spectral}
    }

    \caption{Output-score distributions on BigCloneBench for the learned latent spectrum of \method{} and the fixed AST spectrum, both evaluated without a learned downstream prediction head.}
    \label{fig:rq1_score_distributions}
\end{figure}

A similar pattern appears on the balanced AtCoder benchmark, where accuracy is the relevant measure; the naive baseline is .50, and the untrained fixed-graph spectra remain around chance (.49--.52). Relative to the strongest corresponding fixed-graph configuration, the learned spectrum improves accuracy by .08 with no training, .09 with RF, .11 with LR, and .12 with SNN, reaching .71 with SNN. The complete \method{} further increases accuracy to .81, showing an additional benefit from its learned prediction head. The score distributions in \autoref{fig:rq1_score_distributions} illustrate the same difference under the no-training comparison. The fixed AST spectrum (\subfigref{fig:rq1_score_distributions}{fig:rq1_best_spectral}) leaves clone and non-clone scores almost fully overlapping, whereas the learned latent spectrum (\subfigref{fig:rq1_score_distributions}{fig:rq1_method}) provides a much better separation.

\begin{table}[!htbp]
\centering
\caption{Performance of \method{} and spectral baselines on BigCloneBench and AtCoder. The latent-spectrum rows use the learned spectral representation of \method{} without its original MLP prediction head in \autoref{sec:head}. Parenthesized values report the absolute change from the \emph{Predict All Clone} baseline; $\uparrow$ and $\downarrow$ denote improvement and degradation.}
\label{tab:rq1}

\footnotesize
\setlength{\tabcolsep}{2.0pt}
\renewcommand{\arraystretch}{1.00}

\begin{adjustbox}{max width=\columnwidth}
\begin{tabular}{@{}lcccccccc@{}}
\toprule
& \multicolumn{4}{c}{\textbf{BigCloneBench}}
& \multicolumn{4}{c}{\textbf{AtCoder}} \\
\cmidrule(lr){2-5}
\cmidrule(l){6-9}
\textbf{Method}
& \textbf{P} & \textbf{R} & \textbf{F1} & \textbf{Acc.}
& \textbf{P} & \textbf{R} & \textbf{F1} & \textbf{Acc.} \\
\midrule

\textbf{\method{}}
& .73 & .79 & \textbf{.76\,\gain{.52}} & .93
& .78 & .86 & .82 & \textbf{.81\,\gain{.31}} \\

\midrule

\rowcolor{black!7}
\multicolumn{9}{@{}l}{\textit{Learned latent spectral representation}} \\

\textbf{\method{} Spectrum + No Train}
& .43 & .59 & \textbf{.50\,\gain{.26}} & .84
& .60 & .60 & .60 & \textbf{.60\,\gain{.10}} \\

\textbf{\method{} Spectrum + RF}
& .67 & .64 & \textbf{.65\,\gain{.41}} & .91
& .65 & .82 & .73 & \textbf{.69\,\gain{.19}} \\

\textbf{\method{} Spectrum + LR}
& .61 & .61 & \textbf{.61\,\gain{.37}} & .89
& .62 & .81 & .70 & \textbf{.66\,\gain{.16}} \\

\textbf{\method{} Spectrum + SNN}
& .67 & .67 & \textbf{.67\,\gain{.43}} & .91
& .69 & .78 & .73 & \textbf{.71\,\gain{.21}} \\

\midrule

\rowcolor{black!7}
\multicolumn{9}{@{}l}{\textit{Fixed-graph spectral representations}} \\

AST + No Train
& .14 & .96 & .24\,\samev & .17
& .50 & .66 & .57 & .50\,\samev \\

AST + RF
& .33 & .41 & .37\,\gain{.13} & .81
& .58 & .67 & .62 & \textbf{.60\,\gain{.10}} \\

AST + LR
& .18 & .44 & .26\,\smallgain{.02} & .66
& .52 & .61 & .56 & .52\,\smallgain{.02} \\

AST + SNN
& .35 & .39 & \textbf{.37\,\gain{.13}} & .82
& .56 & .70 & .62 & .57\,\gain{.07} \\

\midrule

CFG + No Train
& .14 & .94 & .24\,\samev & .20
& .51 & .92 & .66 & .52\,\smallgain{.02} \\

CFG + RF
& .25 & .41 & .31\,\smallgain{.07} & .75
& .60 & .65 & .62 & .60\,\gain{.10} \\

CFG + LR
& .21 & .44 & .29\,\smallgain{.05} & .70
& .53 & .76 & .63 & .55\,\gain{.05} \\

CFG + SNN
& .28 & .37 & .32\,\smallgain{.08} & .78
& .57 & .75 & .65 & .59\,\gain{.09} \\

\midrule

DDG + No Train
& .16 & .60 & .25\,\smallgain{.01} & .50
& .50 & .19 & .28 & .50\,\samev \\

DDG + RF
& .25 & .39 & .30\,\smallgain{.06} & .76
& .57 & .64 & .61 & .58\,\gain{.08} \\

DDG + LR
& .20 & .41 & .26\,\smallgain{.02} & .69
& .51 & .76 & .61 & .51\,\smallgain{.01} \\

DDG + SNN
& .29 & .35 & .32\,\smallgain{.08} & .80
& .56 & .69 & .62 & .58\,\gain{.08} \\

\midrule

CPG + No Train
& .14 & .96 & .25\,\smallgain{.01} & .20
& .50 & .88 & .64 & .49\,\smalldrop{.01} \\

CPG + RF
& .33 & .42 & .37\,\gain{.13} & .80
& .58 & .65 & .61 & .59\,\gain{.09} \\

CPG + LR
& .20 & .50 & .29\,\smallgain{.05} & .67
& .52 & .73 & .61 & .53\,\smallgain{.03} \\

CPG + SNN
& .34 & .38 & .36\,\gain{.12} & .81
& .55 & .69 & .62 & .57\,\gain{.07} \\

\midrule

\textbf{Predict All Clone}
& .14 & 1.00 & .24\,\samev & .14
& .50 & 1.00 & .67 & .50\,\samev \\

\bottomrule
\end{tabular}
\end{adjustbox}
\end{table}

\subsection{RQ2: Comparative Effectiveness Against Existing Representations and Clone-Detection Methods}
\label{sec:rq2}
\autoref{tab:rq2_overall} compares \method{} with the baselines described in \autoref{sec:baselines}. The canonical node labels and lexical residuals introduced in \autoref{sec:input_representation} improve performance by +.34 F1 on BigCloneBench, +.30 accuracy on AtCoder, and +.16 accuracy on CodeNet, with lexical residuals providing an additional +.02 accuracy gain on CodeNet. The final \method{} achieves .76 F1 on BigCloneBench, .81 accuracy on AtCoder, and .69 accuracy on CodeNet. On BigCloneBench, it outperforms Deckard (.36 F1) and all implemented GNN baselines, the best of which reaches .35 F1. 

RtvNN and DeepSim reach .93 F1 on BigCloneBench, possibly by exploiting Java-specific lexical and CFG features, the same two models degrade by .228 and .170 under language shift in \autoref{sec:rq3}, whereas \method{} degrades only by .058. Within our shared four-epoch budget (\autoref{sec:settings}), \method{} also continues to improve with training (see \autoref{sec:num_epochs}), so the gap partly reflects the budget rather than a representational ceiling. 

On AtCoder, \method{} is competitive with graph and pre-trained approaches, matching CodeBERT + SNN and remaining close to ASTNN and GraphCodeBERT + SNN in accuracy, while exceeding conventional graph baselines such as GNN-CPG (.56) and DeepSim (.71). On CodeNet, where no external pretraining is used, \method{} achieves higher accuracy than ASTNN (.68), DeepSim (.67), and all GNN-based representations, while approaching stronger pre-trained models such as GraphCodeBERT + SNN (.79).

\begin{table}[!htbp]
\centering
\caption{Test performance of \method{} against competing representations and clone-detection methods on BigCloneBench, AtCoder, and CodeNet. Rows highlighted in green mark the best performing baseline within each family.}
\label{tab:rq2_overall}
\scriptsize

\setlength{\tabcolsep}{1.4pt}
\renewcommand{\arraystretch}{1.0}
\setlength{\aboverulesep}{0.2ex}
\setlength{\belowrulesep}{0.2ex}

\begin{adjustbox}{
    max width=\columnwidth,
    max totalheight=1.0\textheight,
    keepaspectratio,
    center
}
\begin{tabular}{@{}l*{12}{c}@{}}
\toprule
\textbf{Method}
& \multicolumn{4}{c}{\shortstack{\textbf{BigCloneBench}}}
& \multicolumn{4}{c}{\textbf{AtCoder}}
& \multicolumn{4}{c}{\textbf{CodeNet}} \\
\cmidrule(lr){2-5}
\cmidrule(lr){6-9}
\cmidrule(l){10-13}
& \textbf{P} & \textbf{R} & \textbf{F1} & \textbf{Acc.}
& \textbf{P} & \textbf{R} & \textbf{F1} & \textbf{Acc.}
& \textbf{P} & \textbf{R} & \textbf{F1} & \textbf{Acc.} \\
\midrule

\rowcolor{black!7}
\multicolumn{13}{@{}l}{\textit{Our method}}\\[0pt]

\textbf{\method{} (Topo.)}
& .40 & .45 & .42 & .83
& .50 & .69 & .58 & .51
& .52 & .98 & .68 & .53 \\

\textbf{\method{} (Label)}
& .76 & .75 & .76 & .93
& .76 & .90 & .82 & .81
& .61 & .90 & .73 & .67 \\

\rowcolor{green!15}
\textbf{\method{} (Lex.)}
& .73 & .79 & .76 & .93
& .78 & .86 & .82 & .81
& .63 & .89 & .74 & .69 \\

\rowcolor{black!7}
\multicolumn{13}{@{}l}{\textit{Non-graph code baselines}}\\[0pt]

Deckard
& .37 & .36 & .36 & .83
& .52 & .32 & .39 & .51
& .50 & 1.00 & .67 & .50 \\

\rowcolor{green!15}
RtvNN
& .95 & .91 & .93 & .98
& .83 & .87 & .85 & .85
& .68 & .87 & .76 & .73 \\

CDLH
& .93 & .90 & .91 & .98
& .74 & .91 & .82 & .79
& .63 & .86 & .73 & .68 \\

\rowcolor{black!7}
\multicolumn{13}{@{}l}{\textit{Other graph-based learning methods}}\\[0pt]

\rowcolor{green!15}
ASTNN
& .85 & .81 & .83 & .95
& .83 & .90 & .86 & .86
& .62 & .90 & .74 & .68 \\

FA-AST+GGNN
& .94 & .86 & .90 & .97
& .75 & .95 & .84 & .82
& .58 & .90 & .71 & .62 \\

FA-AST+GMN
& .95 & .87 & .91 & .98
& .60 & .66 & .63 & .61
& .57 & .92 & .70 & .62 \\

\rowcolor{black!7}
\multicolumn{13}{@{}l}{\textit{Hybrid graph + code baseline}}\\[0pt]

\rowcolor{green!15}
DeepSim
& .95 & .91 & .93 & .98
& .70 & .75 & .72 & .71
& .61 & .91 & .73 & .67 \\

\rowcolor{black!7}
\multicolumn{13}{@{}l}{\textit{pre-trained code models}}\\[0pt]

CodeBERT + No Train
& .17 & .43 & .24 & .63
& .50 & 1.00 & .67 & .50
& .51 & .98 & .67 & .51 \\

CodeBERT + RF
& .79 & .74 & .77 & .94
& .65 & .59 & .62 & .64
& .60 & .90 & .72 & .65 \\

CodeBERT + SNN
& .82 & .73 & .77 & .94
& .72 & .87 & .79 & .76
& .74 & .88 & .80 & .78 \\

CodeBERT + PCA + RF
& .81 & .70 & .75 & .94
& .67 & .57 & .61 & .64
& .53 & .94 & .68 & .56 \\

CodeBERT + PCA + SNN
& .81 & .82 & .81 & .95
& .72 & .91 & .80 & .78
& .57 & .94 & .71 & .61 \\

GraphCodeBERT + No Train
& .36 & .45 & .40 & .82
& .50 & 1.00 & .67 & .50
& .50 & .99 & .67 & .50 \\

GraphCodeBERT + RF
& .88 & .81 & .84 & .96
& .67 & .56 & .61 & .64
& .59 & .87 & .71 & .64 \\

\rowcolor{green!15}
GraphCodeBERT + SNN
& .90 & .91 & .90 & .97
& .80 & .93 & .86 & .85
& .74 & .90 & .81 & .79 \\

GraphCodeBERT + PCA + RF
& .83 & .82 & .82 & .95
& .70 & .61 & .65 & .67
& .56 & .90 & .69 & .60 \\

GraphCodeBERT + PCA + SNN
& .92 & .86 & .89 & .97
& .73 & .90 & .81 & .79
& .59 & .94 & .73 & .65 \\

\rowcolor{black!7}
\multicolumn{13}{@{}l}{\textit{GNN baselines}}\\[0pt]

GNN-AST
& .21 & .51 & .30 & .68
& .54 & .70 & .61 & .55
& .62 & .69 & .66 & .64 \\

GNN-CFG
& .21 & .46 & .29 & .69
& .59 & .67 & .63 & .60
& .63 & .68 & .65 & .64 \\

GNN-DDG
& .22 & .50 & .31 & .69
& .54 & .52 & .53 & .54
& .63 & .71 & .67 & .65 \\

GNN-CPG
& .27 & .49 & .35 & .75
& .56 & .56 & .56 & .56
& .64 & .62 & .63 & .64 \\

\bottomrule
\end{tabular}
\end{adjustbox}

\end{table}

\subsection{RQ3: Bridge-Assisted Cross-Language Generalization}
\label{sec:rq3}
The central result of this section is robustness. Without training on any cross-language target pairs $Y$--$Y$, every baseline collapses to near-chance on at least four of six language pairs (\autoref{tab:rq3_controls}), while \method{} stays above chance on all six and reaches .71 on Java–Python. Across all 60 bridge paths, it shows the smallest degradation under language shift (.058, versus .112–.228, \autoref{tab:rq3_effects}). The remainder of this section details these results by transfer settings defined in \autoref{sec:protocol}. \autoref{tab:rq3_controls} provides the within-language and endpoint-only controls. \autoref{tab:rq3_bridge_l1} reports length-1 bridge-assisted transfer. \autoref{tab:rq3_bridge_l2} and \autoref{tab:rq3_bridge_l3} extend this analysis to longer bridges and evaluate the effect of intermediate languages and their reinforcement.

\begin{table}[!htbp]
\centering
\caption{Within- and cross-language generalization controls on CodeNet. J, P, C, and S denote Java, Python, C++, and C\#, respectively. The \emph{SL train} column lists the languages whose same-language pairs are used for training; for example, J,P denotes training on J--J and P--P. All values are accuracy.}
\label{tab:rq3_controls}

\footnotesize
\setlength{\tabcolsep}{2.5pt}
\renewcommand{\arraystretch}{1.00}

\resizebox{0.95\columnwidth}{!}{%
\begin{tabular}{@{}llccccc@{}}
\toprule
\textbf{SL train}
& \textbf{Test}
& \textbf{\method{}}
& \textbf{ASTNN}
& \textbf{DeepSim}
& \textbf{RtvNN}
& \textbf{GraphCodeBERT} \\
\midrule

\rowcolor{black!7}
\multicolumn{7}{@{}l}{\textit{Within-language reference}} \\

J & J--J
& .70 & .73 & .77 & .79 & .61 \\

P & P--P
& .73 & .74 & .76 & .79 & .70 \\

C & C--C
& .66 & .60 & .73 & .70 & .64 \\

S & S--S
& .65 & .59 & .74 & .72 & .59 \\

\midrule

\rowcolor{black!7}
\multicolumn{7}{@{}l}{\textit{Cross-language generalization without cross-language supervision}} \\

J,P & J--P
& .71 & .54 & .51 & .49 & .50 \\

J,C & J--C
& .69 & .53 & .70 & .61 & .54 \\

J,S & J--S
& .63 & .54 & .49 & .49 & .60 \\

P,C & P--C
& .55 & .41 & .64 & .56 & .50 \\

P,S & P--S
& .59 & .53 & .66 & .50 & .50 \\

C,S & C--S
& .59 & .47 & .48 & .48 & .50 \\

\bottomrule
\end{tabular}%
}
\end{table}

\begin{table}[!htbp]
\centering
\caption{Length-1 bridge-assisted generalization on CodeNet. Each path $X_1\!\rightarrow Y$ is trained using $X_1$--$X_1$ and $X_1$--$Y$, with no $Y$--$Y$ training pairs. Accuracy is measured on $Y$--$Y$ test pairs.}
\label{tab:rq3_bridge_l1}

\footnotesize
\setlength{\tabcolsep}{3.0pt}
\renewcommand{\arraystretch}{1.2}

\begin{adjustbox}{width=0.95\columnwidth}
\begin{tabular}{@{}lccccc@{}}
\toprule
\textbf{Training path}
& \textbf{\method{}}
& \textbf{ASTNN}
& \textbf{RtvNN}
& \textbf{DeepSim}
& \textbf{GraphCodeBERT} \\
\midrule

\rowcolor{black!7}
\multicolumn{6}{l}{\textit{Target C\# (S)}} \\

J $\rightarrow$ S
& .5007 & .5653 & .5713 & .5000 & .5960 \\
P $\rightarrow$ S
& .5753 & .5040 & .6240 & .6967 & .5120 \\
C $\rightarrow$ S
& .5673 & .4953 & .5687 & .5460 & .5607 \\

\addlinespace[1pt]
\rowcolor{black!7}
\multicolumn{6}{l}{\textit{Target Java (J)}} \\

P $\rightarrow$ J
& .6364 & .7051 & .7298 & .6618 & .5033 \\
C $\rightarrow$ J
& .6878 & .6511 & .7191 & .7171 & .6227 \\
S $\rightarrow$ J
& .5917 & .5284 & .6664 & .4570 & .6347 \\

\addlinespace[1pt]
\rowcolor{black!7}
\multicolumn{6}{l}{\textit{Target Python (P)}} \\

J $\rightarrow$ P
& .6083 & .7418 & .7029 & .5791 & .5707 \\
C $\rightarrow$ P
& .6144 & .6130 & .6841 & .6640 & .5540 \\
S $\rightarrow$ P
& .5446 & .5399 & .6875 & .7101 & .5627 \\

\addlinespace[1pt]
\rowcolor{black!7}
\multicolumn{6}{l}{\textit{Target C++ (C)}} \\

J $\rightarrow$ C
& .6573 & .5547 & .6860 & .7254 & .6287 \\
P $\rightarrow$ C
& .6080 & .5780 & .6513 & .6711 & .5093 \\
S $\rightarrow$ C
& .5627 & .4753 & .5800 & .6269 & .6113 \\

\bottomrule
\end{tabular}
\end{adjustbox}
\end{table}

\begin{table}[!htbp]
\centering
\caption{Length-2 bridge-assisted generalization on CodeNet. A path $X_1\!\rightarrow X_2\!\rightarrow Y$ uses $X_1$--$X_1$, $X_1$--$X_2$, and $X_2$--$Y$ as mandatory training configurations. Each cell reports accuracy as $\varnothing$ / $X_2$--$X_2$, where the second value additionally includes same-language reinforcement for the intermediate language. Evaluation is on $Y$--$Y$ test pairs.}
\label{tab:rq3_bridge_l2}

\scriptsize
\setlength{\tabcolsep}{6.4pt}
\renewcommand{\arraystretch}{1.2}

\begin{adjustbox}{width=0.95\columnwidth}
\begin{tabular}{lccccc}
\toprule
\textbf{Training path}
& \textbf{\method{}}
& \textbf{ASTNN}
& \textbf{RtvNN}
& \textbf{DeepSim}
& \textbf{GraphCodeBERT} \\
\midrule

\rowcolor{black!7}
\multicolumn{6}{l}{\textit{Target C\# (S)}} \\

J $\rightarrow$ P $\rightarrow$ S
& .5993/.6013 & .5040/.5667 & .5653/.6187 & .5927/.6700 & .5400/.5707 \\
J $\rightarrow$ C $\rightarrow$ S
& .5940/.5740 & .5007/.5007 & .5920/.5220 & .5620/.5227 & .5787/.6087 \\
P $\rightarrow$ J $\rightarrow$ S
& .6307/.5927 & .5093/.5020 & .5893/.5927 & .6207/.5960 & .5507/.6053 \\
P $\rightarrow$ C $\rightarrow$ S
& .6053/.5467 & .4947/.4980 & .5480/.5760 & .6327/.6053 & .5287/.5520 \\
C $\rightarrow$ J $\rightarrow$ S
& .5947/.5640 & .5100/.5033 & .5613/.5567 & .6047/.5720 & .5900/.6447 \\
C $\rightarrow$ P $\rightarrow$ S
& .5920/.5987 & .5087/.5247 & .5627/.6287 & .6687/.6473 & .5253/.5440 \\

\addlinespace[1pt]
\rowcolor{black!7}
\multicolumn{6}{l}{\textit{Target Java (J)}} \\

P $\rightarrow$ C $\rightarrow$ J
& .6905/.6785 & .6451/.6538 & .6871/.6905 & .6611/.6511 & .5547/.5960 \\
P $\rightarrow$ S $\rightarrow$ J
& .6164/.6324 & .5664/.5997 & .6324/.6564 & .6584/.6204 & .6033/.6527 \\
C $\rightarrow$ P $\rightarrow$ J
& .6744/.6811 & .6211/.6751 & .6891/.6865 & .6344/.6384 & .5180/.5540 \\
C $\rightarrow$ S $\rightarrow$ J
& .6531/.7031 & .5991/.6084 & .6785/.7011 & .6137/.6284 & .6547/.6887 \\
S $\rightarrow$ P $\rightarrow$ J
& .6224/.6771 & .5984/.6631 & .6498/.6798 & .6638/.5824 & .5253/.5607 \\
S $\rightarrow$ C $\rightarrow$ J
& .6478/.6885 & .5564/.6438 & .6698/.7038 & .6771/.6664 & .6220/.6813 \\

\addlinespace[1pt]
\rowcolor{black!7}
\multicolumn{6}{l}{\textit{Target Python (P)}} \\

J $\rightarrow$ C $\rightarrow$ P
& .6848/.6821 & .6559/.6740 & .7190/.6935 & .6606/.6273 & .6393/.6900 \\
J $\rightarrow$ S $\rightarrow$ P
& .6378/.6908 & .6284/.6197 & .6901/.7163 & .5995/.6836 & .6607/.7180 \\
C $\rightarrow$ J $\rightarrow$ P
& .6190/.6164 & .6808/.7183 & .6633/.7156 & .6551/.6382 & .6587/.7200 \\
C $\rightarrow$ S $\rightarrow$ P
& .6244/.6351 & .5808/.5996 & .6097/.7103 & .6544/.6599 & .6660/.6880 \\
S $\rightarrow$ J $\rightarrow$ P
& .6492/.6526 & .6546/.7223 & .6486/.7210 & .6001/.6415 & .6940/.7187 \\
S $\rightarrow$ C $\rightarrow$ P
& .6392/.6432 & .5687/.6566 & .6640/.7062 & .6741/.6653 & .6467/.7100 \\

\addlinespace[1pt]
\rowcolor{black!7}
\multicolumn{6}{l}{\textit{Target C++ (C)}} \\

J $\rightarrow$ P $\rightarrow$ C
& .6587/.6293 & .5813/.5953 & .6480/.6767 & .6417/.6490 & .5200/.5540 \\
J $\rightarrow$ S $\rightarrow$ C
& .6407/.6427 & .5687/.5307 & .6387/.6427 & .6437/.6189 & .6487/.6940 \\
P $\rightarrow$ J $\rightarrow$ C
& .6147/.6960 & .5793/.5813 & .6607/.6553 & .6872/.6524 & .5747/.6367 \\
P $\rightarrow$ S $\rightarrow$ C
& .6253/.5573 & .5540/.5213 & .5913/.6147 & .6048/.6222 & .6040/.6213 \\
S $\rightarrow$ J $\rightarrow$ C
& .6420/.6473 & .5020/.5660 & .6533/.6773 & .6296/.6370 & .6400/.6953 \\
S $\rightarrow$ P $\rightarrow$ C
& .5853/.6040 & .5580/.5793 & .6173/.6487 & .6088/.6236 & .5287/.5307 \\

\bottomrule
\end{tabular}
\end{adjustbox}
\end{table}

\begin{table*}[!htbp]
\centering
\caption{Length-3 bridge-assisted generalization on CodeNet. A path $X_1\!\rightarrow X_2\!\rightarrow X_3\!\rightarrow Y$ uses $X_1$--$X_1$, $X_1$--$X_2$, $X_2$--$X_3$, and $X_3$--$Y$ as mandatory training configurations, with no $Y$--$Y$ pairs. Each cell reports accuracy as $\varnothing$ / $X_2$--$X_2$ / $X_3$--$X_3$ / $X_2$--$X_2$+$X_3$--$X_3$, where the later values additionally include same-language reinforcement for the intermediate languages. Evaluation is on $Y$--$Y$ test pairs.}
\label{tab:rq3_bridge_l3}

\scriptsize
\setlength{\tabcolsep}{8.0pt}
\renewcommand{\arraystretch}{1.2}
\setlength{\aboverulesep}{0.15ex}
\setlength{\belowrulesep}{0.15ex}

\begin{adjustbox}{width=0.95\textwidth}
\begin{tabular}{lccccc}
\toprule
\textbf{Training path}
& \textbf{\method{}}
& \textbf{ASTNN}
& \textbf{RtvNN}
& \textbf{DeepSim}
& \textbf{GraphCodeBERT} \\
\midrule

\rowcolor{black!7}
\multicolumn{6}{l}{\textit{Target C\# (S)}} \\

J $\rightarrow$ P $\rightarrow$ C $\rightarrow$ S
& .5393/.5993/.5720/.5987
& .5020/.4993/.5007/.5053
& .5453/.5720/.5453/.5653
& .5973/.6553/.5780/.5993
& .5727/.6067/.5867/.6240 \\

J $\rightarrow$ C $\rightarrow$ P $\rightarrow$ S
& .5313/.5867/.5833/.5907
& .5007/.5133/.5033/.4887
& .5460/.5500/.5787/.6013
& .6100/.6413/.6500/.6873
& .5500/.5827/.5813/.6120 \\

P $\rightarrow$ J $\rightarrow$ C $\rightarrow$ S
& .5127/.5320/.5160/.5673
& .4940/.4973/.4973/.4953
& .5420/.5400/.5413/.5467
& .5920/.5580/.5773/.5733
& .5673/.5953/.5907/.6180 \\

P $\rightarrow$ C $\rightarrow$ J $\rightarrow$ S
& .5587/.5753/.5540/.5673
& .5087/.4947/.5180/.5500
& .5573/.5333/.5600/.5340
& .6200/.5933/.5667/.5833
& .5613/.6240/.5820/.6360 \\

C $\rightarrow$ J $\rightarrow$ P $\rightarrow$ S
& .5773/.6240/.5533/.5907
& .5087/.5060/.5220/.5813
& .5593/.5713/.5893/.5753
& .6373/.5860/.6620/.6480
& .5520/.5980/.6007/.6413 \\

C $\rightarrow$ P $\rightarrow$ J $\rightarrow$ S
& .5487/.6047/.5607/.5640
& .4980/.5080/.5680/.5140
& .5813/.5600/.5687/.6180
& .5647/.5947/.6180/.6260
& .5893/.6187/.6307/.6633 \\

\addlinespace[1pt]
\rowcolor{black!7}
\multicolumn{6}{l}{\textit{Target Java (J)}} \\

P $\rightarrow$ C $\rightarrow$ S $\rightarrow$ J
& .5757/.6464/.6077/.6418
& .5837/.6244/.6071/.5777
& .6684/.6518/.5911/.6491
& .6678/.6271/.6184/.6331
& .5820/.6580/.6380/.6820 \\

P $\rightarrow$ S $\rightarrow$ C $\rightarrow$ J
& .6471/.6711/.6611/.6604
& .5777/.6271/.5991/.6211
& .6611/.6744/.6191/.6791
& .6444/.6231/.6711/.6624
& .6453/.6553/.6467/.7000 \\

C $\rightarrow$ P $\rightarrow$ S $\rightarrow$ J
& .5804/.6518/.5884/.6498
& .5777/.5984/.6111/.6044
& .6404/.6845/.6458/.6818
& .6111/.6131/.6124/.6291
& .6027/.6560/.6480/.6840 \\

C $\rightarrow$ S $\rightarrow$ P $\rightarrow$ J
& .6564/.6631/.6037/.6451
& .6251/.6211/.6771/.6718
& .6785/.6518/.7085/.6931
& .6217/.5670/.6258/.6104
& .5747/.6133/.6013/.6660 \\

S $\rightarrow$ P $\rightarrow$ C $\rightarrow$ J
& .6404/.6404/.5550/.6758
& .5664/.6271/.6351/.6411
& .7171/.6758/.6798/.6658
& .6404/.6017/.5944/.6684
& .5520/.6187/.6180/.6273 \\

S $\rightarrow$ C $\rightarrow$ P $\rightarrow$ J
& .6004/.6338/.6157/.6624
& .6431/.6584/.6798/.6604
& .6744/.6891/.7058/.6938
& .6251/.6177/.6151/.6011
& .5853/.6507/.6200/.6700 \\

\addlinespace[1pt]
\rowcolor{black!7}
\multicolumn{6}{l}{\textit{Target Python (P)}} \\

J $\rightarrow$ C $\rightarrow$ S $\rightarrow$ P
& .6848/.6774/.6861/.6311
& .6714/.6117/.5842/.6559
& .6459/.5848/.7324/.7398
& .6646/.6741/.6667/.7040
& .7080/.7287/.7020/.7527 \\

J $\rightarrow$ S $\rightarrow$ C $\rightarrow$ P
& .5211/.6432/.5547/.6338
& .6284/.6304/.6613/.6512
& .6767/.7009/.7103/.6982
& .6599/.6110/.6531/.6470
& .6867/.7347/.6980/.7493 \\

C $\rightarrow$ J $\rightarrow$ S $\rightarrow$ P
& .5734/.7049/.6023/.6747
& .6177/.6901/.6123/.6600
& .5989/.5842/.7150/.7391
& .6640/.6219/.6395/.5737
& .6753/.7307/.7267/.7387 \\

C $\rightarrow$ S $\rightarrow$ J $\rightarrow$ P
& .6144/.6593/.5560/.6177
& .6512/.6365/.7136/.6875
& .6908/.6868/.6378/.7029
& .6341/.6239/.6687/.6517
& .6880/.7273/.7333/.7507 \\

S $\rightarrow$ J $\rightarrow$ C $\rightarrow$ P
& .6372/.6660/.5755/.6224
& .5996/.6539/.6848/.6667
& .7042/.6989/.6640/.7384
& .5913/.6191/.6314/.6619
& .6853/.7287/.7353/.7540 \\

S $\rightarrow$ C $\rightarrow$ J $\rightarrow$ P
& .6318/.6848/.6653/.6962
& .6687/.6754/.7210/.6982
& .6479/.6968/.7217/.7337
& .4657/.6551/.6667/.6449
& .6847/.7213/.6927/.7413 \\

\addlinespace[1pt]
\rowcolor{black!7}
\multicolumn{6}{l}{\textit{Target C++ (C)}} \\

J $\rightarrow$ P $\rightarrow$ S $\rightarrow$ C
& .6333/.6633/.6267/.6193
& .5587/.5880/.5380/.5440
& .6113/.6180/.6040/.6167
& .6263/.6517/.4943/.5948
& .5887/.6287/.6307/.6700 \\

J $\rightarrow$ S $\rightarrow$ P $\rightarrow$ C
& .5720/.6447/.5727/.6493
& .5673/.5420/.5980/.5793
& .6393/.6520/.6500/.6800
& .6035/.5948/.6155/.6229
& .5880/.6347/.6120/.6713 \\

P $\rightarrow$ J $\rightarrow$ S $\rightarrow$ C
& .5527/.6740/.5533/.6000
& .5547/.5700/.5533/.5433
& .6067/.6313/.6020/.6280
& .6048/.6457/.6330/.6316
& .6380/.6767/.6707/.7093 \\

P $\rightarrow$ S $\rightarrow$ J $\rightarrow$ C
& .5433/.6267/.5873/.6487
& .5773/.5627/.5360/.6040
& .6640/.6140/.6673/.6740
& .6658/.6557/.6537/.6155
& .6413/.6900/.6733/.6947 \\

S $\rightarrow$ J $\rightarrow$ P $\rightarrow$ C
& .5960/.6573/.5933/.6253
& .5400/.5773/.5340/.5933
& .6380/.6633/.6620/.6707
& .6082/.6222/.6336/.6149
& .5453/.6207/.6167/.6833 \\

S $\rightarrow$ P $\rightarrow$ J $\rightarrow$ C
& .5867/.6113/.6087/.6387
& .5600/.5960/.5673/.5953
& .6547/.6673/.6580/.6653
& .6356/.6450/.6236/.6517
& .5440/.6400/.6513/.6880 \\

\bottomrule
\end{tabular}
\end{adjustbox}
\end{table*}

The statistical analyses in \autoref{tab:rq3_significance} and \autoref{tab:rq3_effects} provide a quantitative summary of cross-language transfer behavior. Over Java--Python test pairs within the cross-language generalization settings in \autoref{tab:rq3_controls}, baselines ASTNN, DeepSim, RtvNN, and GraphCodeBERT obtain .54, .51, .49, and .50 accuracy, i.e. chance level, while \method{} reaches .71 accuracy. Across all 60 bridge paths, \method{} achieves higher average accuracy than ASTNN by $.033$ and GraphCodeBERT by $.011$, while remaining below RtvNN only by $.031$ and DeepSim by $.021$. The difference with GraphCodeBERT is not statistically significant, whereas the differences with the other three baselines are. \method{} also exhibits the smallest degradation under language shift, decreasing by $.058$ compared with $.112$ for GraphCodeBERT, $.162$ for ASTNN, $.170$ for DeepSim, and $.228$ for RtvNN. Bridge length has a significant overall effect ($p=.0009$), with length-2 bridges outperforming length-3 bridges by $.043$ ($p=.007$), while the remaining pairwise differences are not significant. Intermediate-language reinforcement consistently improves transfer, providing gains of $.017$ for length-2 bridges and up to $.036$ when both intermediate languages are reinforced in length-3 bridges ($p<.001$). The choice of target language (Y) also has a significant effect ($p=.0006$), with C\# being the most challenging target, particularly compared with Java ($-.066$, $p=.001$) and Python ($-.050$, $p=.009$). In contrast, the source language has no significant effect ($p=.862$).

\begin{table}[!htbp]
\centering
\caption{Factors affecting bridge-assisted transfer for \method{}. Bridge length is tested with Friedman across the three lengths and Wilcoxon for the pairwise contrasts; target and source language with Kruskal--Wallis and Mann--Whitney; reinforcement with Wilcoxon paired per path. All $p$ values are Holm-corrected within their family.}
\label{tab:rq3_effects}

\footnotesize
\setlength{\tabcolsep}{4.5pt}
\renewcommand{\arraystretch}{0.9}

\begin{adjustbox}{width=0.9\columnwidth}
\begin{tabular}{llrrl}
\toprule
\textbf{Factor} & \textbf{Contrast} & \textbf{$\Delta$} & \textbf{$p$} & \textbf{Effect} \\
\midrule
\rowcolor{black!7}
\multicolumn{5}{l}{\textit{Bridge length (Friedman $\chi^2=14.00$, $p=0.0009$)}} \\
 & L1 vs L2 & $-0.035$ & $.130$ & large \\
 & L1 vs L3 & $+0.008$ & $1.000$ & negligible \\
 & L2 vs L3 & $+0.043$ & $.007$ & large \\
\addlinespace[2pt]
\rowcolor{black!7}
\multicolumn{5}{l}{\textit{Target language (Kruskal--Wallis $H=17.50$, $p=0.0006$)}} \\
 & C\# vs C++ & $-0.037$ & $.081$ & large \\
 & C\# vs Java & $-0.066$ & $.001$ & large \\
 & C\# vs Python & $-0.050$ & $.009$ & large \\
 & C++ vs Java & $-0.029$ & $.186$ & medium \\
 & C++ vs Python & $-0.014$ & $.561$ & small \\
 & Java vs Python & $+0.016$ & $.561$ & small \\
\addlinespace[2pt]
\rowcolor{black!7}
\multicolumn{5}{l}{\textit{Bridge source language (Kruskal--Wallis $H=0.75$, $p=0.862$ --- no difference)}} \\
 & C++ & $0.612$ & --- & $\pm0.041$ \\
 & C\# & $0.612$ & --- & $\pm0.033$ \\
 & Java & $0.604$ & --- & $\pm0.060$ \\
 & Python & $0.600$ & --- & $\pm0.046$ \\
\addlinespace[2pt]
\rowcolor{black!7}
\multicolumn{5}{l}{\textit{Intermediate-language reinforcement}} \\
 & L2, $X_2$--$X_2$ & $+0.017$ & $<.001$ & small \\
 & L3, $X_2$--$X_2$ & $+0.023$ & $<.001$ & small \\
 & L3, $X_2$--$X_2$ + $X_3$--$X_3$ & $+0.036$ & $<.001$ & medium \\
 & L3, $X_3$--$X_3$ & $+0.014$ & $<.001$ & negligible \\
\addlinespace[2pt]
\rowcolor{black!7}
\multicolumn{5}{l}{\textit{Degradation under language shift (\autoref{tab:rq3_controls})}} \\
 & SPECTRA-Siam & $0.058$ & --- & 0.685 $\rightarrow$ 0.627 \\
 & GraphCodeBERT & $0.112$ & --- & 0.635 $\rightarrow$ 0.523 \\
 & ASTNN & $0.162$ & --- & 0.665 $\rightarrow$ 0.503 \\
 & DeepSim & $0.170$ & --- & 0.750 $\rightarrow$ 0.580 \\
 & RtvNN & $0.228$ & --- & 0.750 $\rightarrow$ 0.522 \\
\bottomrule
\end{tabular}
\end{adjustbox}
\end{table}

\begin{table}[!htbp]
\centering
\caption{Statistical comparison of \method{} against each baseline on the bridge-transfer paths of \autoref{tab:rq3_bridge_l1}--\autoref{tab:rq3_bridge_l3}. Every path is evaluated by all five methods, so comparisons are paired on the path and tested with a Wilcoxon signed-rank test; $p$ values are Holm-corrected within each family. $\delta$ is Cliff's delta. Positive differences favor \method{}.}
\label{tab:rq3_significance}

\footnotesize
\setlength{\tabcolsep}{4.5pt}
\renewcommand{\arraystretch}{0.9}

\begin{adjustbox}{width=0.9\columnwidth}
\begin{tabular}{lrrrrl}
\toprule
\textbf{Baseline} & \textbf{$\Delta$acc.} & \textbf{W/L} & \textbf{$p$} & \textbf{$\delta$} & \textbf{Effect} \\
\midrule
\rowcolor{black!7}
\multicolumn{6}{l}{\textit{All 60 paths}} \\
ASTNN & $+0.033$ & 47/13 & $<.001$ & $+0.35$ & medium \\
RtvNN & $-0.031$ & 16/44 & $<.001$ & $-0.35$ & medium \\
DeepSim & $-0.021$ & 22/37 & $.006$ & $-0.29$ & small \\
GraphCodeBERT & $+0.011$ & 33/27 & $.213$ & $+0.14$ & negligible \\
\addlinespace[2pt]
\rowcolor{black!7}
\multicolumn{6}{l}{\textit{Length-1 paths ($n=12$)}} \\
ASTNN & $+0.017$ & 9/3 & $.934$ & $+0.24$ & small \\
RtvNN & $-0.060$ & 0/12 & $.005$ & $-0.56$ & large \\
DeepSim & $-0.033$ & 4/8 & $.752$ & $-0.35$ & medium \\
GraphCodeBERT & $+0.024$ & 8/4 & $.934$ & $+0.24$ & small \\

\addlinespace[2pt]
\rowcolor{black!7}
\multicolumn{6}{l}{\textit{Length-2 paths ($n=24$)}} \\
ASTNN & $+0.059$ & 22/2 & $<.001$ & $+0.62$ & large \\
RtvNN & $-0.004$ & 12/12 & $1.000$ & $-0.11$ & negligible \\
DeepSim & $-0.004$ & 12/12 & $1.000$ & $-0.12$ & negligible \\
GraphCodeBERT & $+0.036$ & 17/7 & $.059$ & $+0.34$ & medium \\

\addlinespace[2pt]
\rowcolor{black!7}
\multicolumn{6}{l}{\textit{Length-3 paths ($n=24$)}} \\
ASTNN & $+0.014$ & 16/8 & $.752$ & $+0.15$ & negligible \\
RtvNN & $-0.043$ & 4/20 & $.001$ & $-0.47$ & medium \\
DeepSim & $-0.031$ & 6/17 & $.078$ & $-0.44$ & medium \\
GraphCodeBERT & $-0.021$ & 8/16 & $.752$ & $-0.22$ & small \\
\bottomrule
\end{tabular}
\end{adjustbox}
\end{table}

\section{Discussion}
\label{sec:discussion}

\subsection{\method{} Sensitivity to Hyperparameters}
\label{sec:hyperparameter_sensitivity}
\begin{figure}[!htbp]
    \centering
    \includegraphics[
        width=0.95\columnwidth,
        height= 0.75\textheight,
        keepaspectratio,
        trim=5 0 0 0,
        clip,
    ]{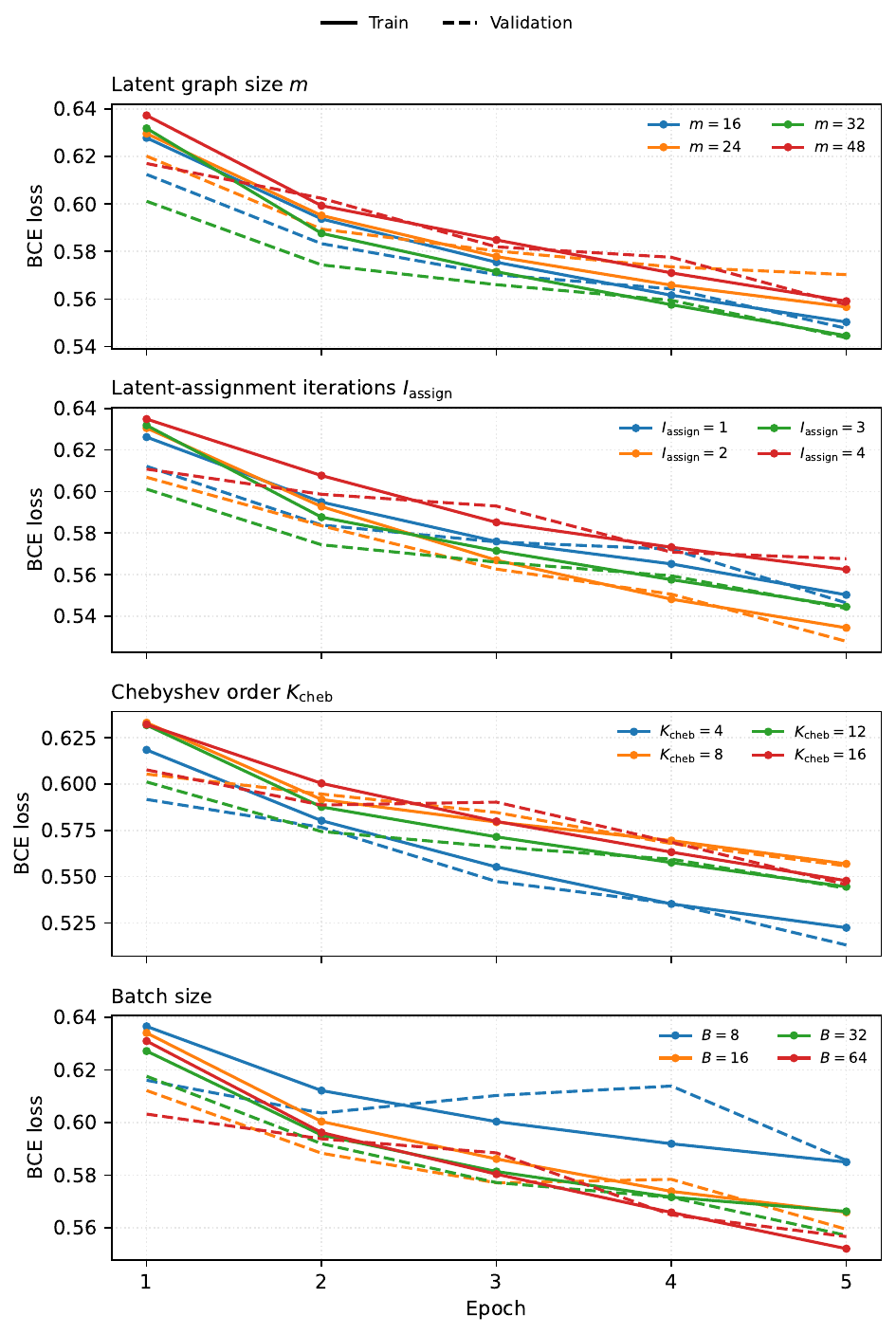}
    \caption{Sensitivity of \method{} to four hyperparameters on CodeNet. From top to bottom, the panels vary latent graph size $m$, latent-assignment iterations $I_{\mathrm{assign}}$, Chebyshev order $K_{\mathrm{cheb}}$, and batch size $B$. Each panel reports training and validation BCE loss over the first five epochs while varying one hyperparameter and keeping all others fixed to the canonical configuration in \autoref{tab:spectra_settings}. Solid and dashed lines denote training and validation loss, respectively.}
    \label{fig:hyperparameter_sensitivity}
\end{figure}

We examine the sensitivity of \method{} on CodeNet by varying one hyperparameter at a time while keeping the remaining settings fixed to the canonical configuration in \autoref{tab:spectra_settings}. The resulting optimization trajectories are shown in \autoref{fig:hyperparameter_sensitivity}, and the corresponding final epoch losses and test accuracies are reported in \autoref{tab:d1_sensitivity}. For the latent graph size, $m=32$ gives consistently low validation loss, while increasing the capacity to $m=48$ provides no clear benefit. For latent assignment, two to three iterations are sufficient, and additional refinement does not improve validation loss. The Chebyshev-order experiment shows that increasing spectral-filter order is not uniformly beneficial over the first five epochs; lower orders converge quickly, while $K_{\mathrm{cheb}}=12$ remains within the same general range. Batch size mainly affects the optimization trajectory: $B=8$ is less stable, whereas $B\in\{16,32,64\}$ gives similar and steadily decreasing validation loss. The canonical configuration provides a stable operating point, though batch size $B$ and the number of assignment iterations $I_{\text{assign}}$ each shift test accuracy by three to four points.

\begin{table}[!htbp]
\centering
\caption{Final-epoch training and validation BCE loss, and test accuracy, for each hyperparameter setting in \autoref{fig:hyperparameter_sensitivity}. All runs use CodeNet with five epochs per arm. Bold marks the best value per block by test accuracy.}
\label{tab:d1_sensitivity}

\footnotesize
\setlength{\tabcolsep}{4.2pt}
\renewcommand{\arraystretch}{1.1}

\begin{tabular}{|c|c|c|c|c|}
\hline
\textbf{Hyperparameter} &
\textbf{Value} &
\textbf{Train BCE} &
\textbf{Valid BCE} &
\textbf{Test Acc.} \\
\hline

\multirow{4}{*}{$m$}
& 16 & 0.5503 & 0.5476 & 0.6960 \\
& 24 & 0.5566 & 0.5703 & 0.6850 \\
& \textbf{32} & \textbf{0.5445} & \textbf{0.5437} & \textbf{0.7029} \\
& 48 & 0.5591 & 0.5576 & 0.6838 \\
\hline

\multirow{4}{*}{$I_{\mathrm{assign}}$}
& 1 & 0.5503 & 0.5464 & 0.7012 \\
& \textbf{2} & \textbf{0.5344} & \textbf{0.5279} & \textbf{0.7056} \\
& 3 & 0.5445 & 0.5437 & 0.7029 \\
& 4 & 0.5625 & 0.5676 & 0.6746 \\
\hline

\multirow{4}{*}{$K_{\mathrm{cheb}}$}
& \textbf{4} & \textbf{0.5224} & \textbf{0.5131} & \textbf{0.7273} \\
& 8 & 0.5569 & 0.5558 & 0.6818 \\
& 12 & 0.5445 & 0.5437 & 0.7029 \\
& 16 & 0.5478 & 0.5457 & 0.6980 \\
\hline

\multirow{4}{*}{$B$}
& 8 & 0.5850 & 0.5856 & 0.6417 \\
& 16 & 0.5659 & 0.5594 & 0.6740 \\
& \textbf{32} & \textbf{0.5662} & \textbf{0.5571} & \textbf{0.6867} \\
& 64 & 0.5520 & 0.5566 & 0.6834 \\
\hline

\end{tabular}
\end{table}

\subsection{Number of Epochs}
\label{sec:num_epochs}
\begin{figure*}[!htbp]
    \centering
    
    \subfloat[Validation loss and training objective.%
    \label{fig:d2_loss_with_train}]
    {
        \includegraphics[
            width=0.95\textwidth
        ]{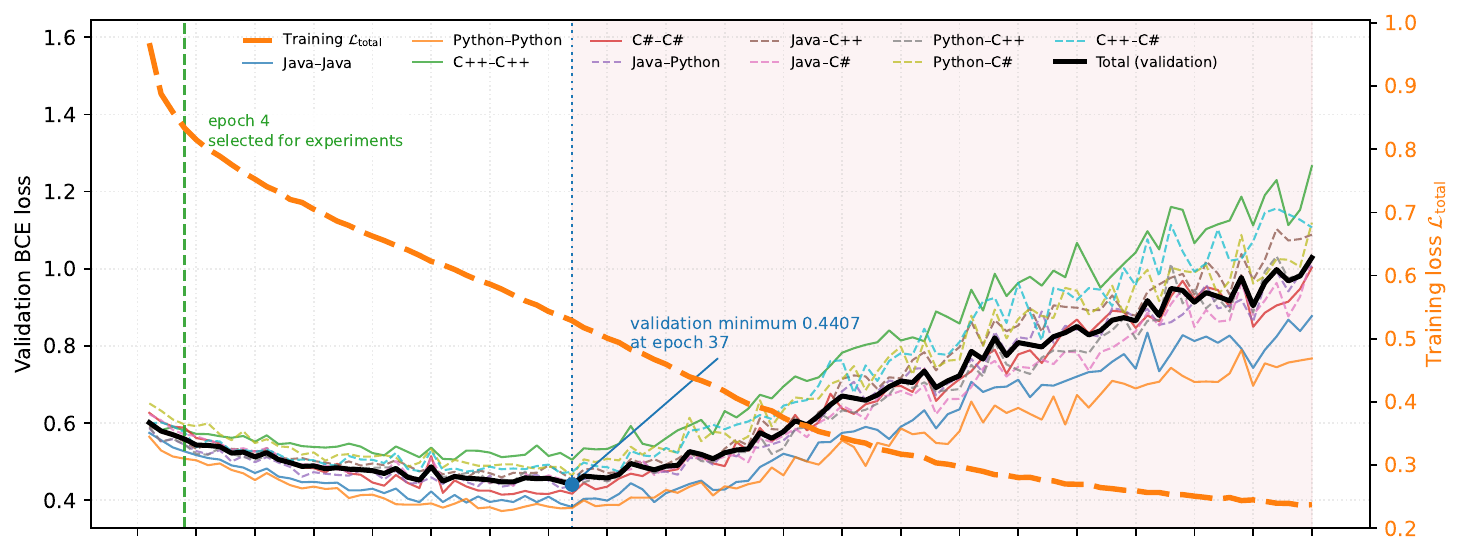}
    }
    
    \vspace{0.5em}
    
    \subfloat[Test accuracy and decision threshold.%
    \label{fig:d2_accuracy_with_threshold}]
    {
        \includegraphics[
            width=0.95\textwidth
        ]{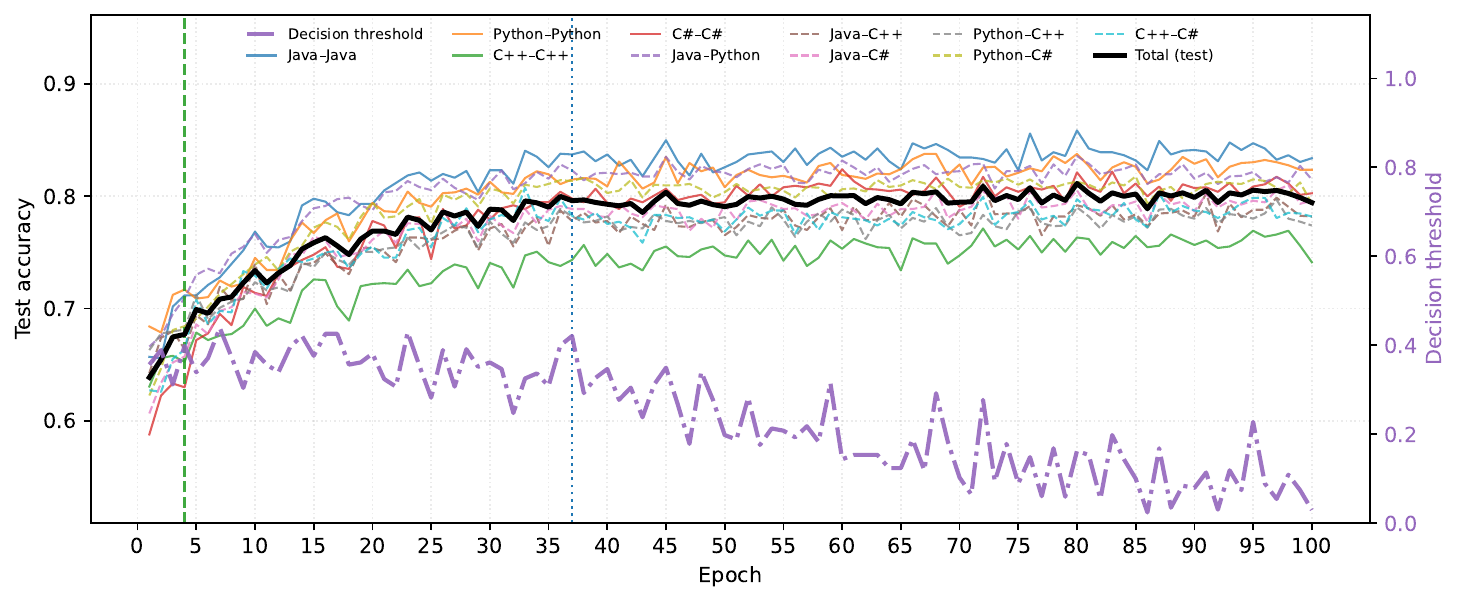}
    }

    \caption{Training dynamics of \method{} on CodeNet over 100 epochs across the ten language configurations (a) Validation BCE loss and the training objective $\mathcal{L}_{\mathrm{total}}$. (b) Test accuracy and the validation-selected decision threshold.}
    \label{fig:d2_training_dynamics}
\end{figure*}

We train \method{} on CodeNet for 100 epochs and evaluate the same model on the validation and test splits after each epoch, reporting both the overall performance and the ten language configurations. As shown in Figure\autoref{fig:d2_loss_with_train}, the training objective $\mathcal{L}_{\mathrm{total}}$ from \autoref{sec:training_objective} (\autoref{eq:complete_loss}) decreases continuously from $.968$ to $.237$. In contrast, the validation cross-entropy reaches its minimum of $.4407$ at epoch $37$ and then increases to $1.0288$ by epoch $100$. All ten language configurations show the same trend, with their minima occurring between epochs $30$ and $37$.

Nevertheless, accuracy continues to improve even after the validation loss starts increasing, reaching $.8114$ at epoch $80$ before slightly decreasing to $.7944$ at epoch $100$ as shown in Figure\autoref{fig:d2_accuracy_with_threshold}. The selected decision threshold also decreases during this period, from $.42$ at epoch $37$ to $.03$ at epoch $100$. Its correlation with validation cross-entropy is $-.892$. We can see that later epochs produce less calibrated probabilities, while the classification accuracy remains strong due to threshold adjustment over validation set.

These results show that the four-epoch training budget used throughout \autoref{sec:exp_results} is conservative. The total accuracy improves from $.6770$ at epoch $4$ to $0.7886$ at epoch $30$ and $.8114$ at epoch $80$, while both same-language and cross-language configurations benefit from longer training. Therefore, the results in \autoref{sec:exp_results} represent a lower-bound estimate of \method{} under the fixed training budget.

\begin{table}[!htbp]
\centering
\caption{Test accuracy of \method{} on CodeNet after selected epochs, reported separately for same-language and cross-language configurations. \emph{Total} denotes accuracy over the complete test set. Bold marks the best value in each column; the highlighted rows correspond to the four-epoch budget of \autoref{sec:exp_results} ($4$), the validation-loss minimum ($37$), and the accuracy peak ($80$).}
\label{tab:d2_epochs}

\footnotesize
\setlength{\tabcolsep}{1.6pt}
\renewcommand{\arraystretch}{0.95}

\newcolumntype{Y}{>{\centering\arraybackslash}X}

\begin{tabularx}{0.95\columnwidth}{|c|Y|Y|Y|Y|Y|}
\hline
\rule{0pt}{2.4ex}\textbf{Ep.} &
\textbf{Total} &
\textbf{J--J} &
\textbf{P--P} &
\textbf{C++--C++} &
\textbf{C\#--C\#} \\
\hline
\rule{0pt}{2.8ex}1 & .6388 & .6571 & .6841 & .6307 & .5880 \\
2 & .6547 & .6558 & .6787 & .6560 & .6227 \\
3 & .6748 & .7018 & .7123 & .6580 & .6333 \\
\rowcolor{black!7}
4 & .6770 & .7118 & .7170 & .6527 & .6300 \\
5 & .6992 & .7118 & .7089 & .6787 & .6720 \\
10 & .7339 & .7685 & .7451 & .7000 & .7140 \\
20 & .7690 & .7932 & .7941 & .7220 & .7780 \\
30 & .7886 & .8232 & .8129 & .7407 & .7793 \\
\rowcolor{black!7}
37 & .7961 & .8372 & .8149 & .7433 & .7980 \\
40 & .7929 & .8372 & .8089 & .7487 & .7940 \\
50 & .7908 & .8259 & .8082 & .7473 & .7947 \\
60 & .8003 & .8352 & .8189 & .7533 & \textbf{.8240} \\
70 & .7947 & .8346 & .8283 & .7473 & .7907 \\
\rowcolor{black!7}
80 & \textbf{.8114} & \textbf{.8586} & \textbf{.8377} & \textbf{.7633} & .8213 \\
90 & .7990 & .8412 & .8290 & .7567 & .8073 \\
100 & .7944 & .8339 & .8236 & .7413 & .8027 \\
\hline
\end{tabularx}

\vspace{-0.4mm}

\begin{tabularx}{0.95\columnwidth}{|c|Y|Y|Y|Y|Y|Y|}
\hline
\rule{0pt}{2.4ex}\textbf{Ep.} &
\textbf{J--P} &
\textbf{J--C++} &
\textbf{J--C\#} &
\textbf{P--C++} &
\textbf{P--C\#} &
\textbf{C++--C\#} \\
\hline
\rule{0pt}{2.8ex}1 & .6662 & .6427 & .6067 & .6631 & .6227 & .6273 \\
2 & .6756 & .6740 & .6340 & .6779 & .6468 & .6260 \\
3 & .6952 & .6800 & .6520 & .6799 & .6810 & .6553 \\
\rowcolor{black!7}
4 & .7093 & .6640 & .6573 & .6812 & .6843 & .6633 \\
5 & .7301 & .6940 & .6860 & .7122 & .6910 & .7080 \\
10 & .7665 & .7400 & .7133 & .7236 & .7393 & .7293 \\
20 & .7873 & .7540 & .7613 & .7505 & .7949 & .7553 \\
30 & .8122 & .7713 & .7827 & .7774 & .8070 & .7793 \\
\rowcolor{black!7}
37 & .8230 & .7780 & .7887 & .7821 & \textbf{.8137} & .7827 \\
40 & .8210 & .7780 & .7813 & .7821 & .8029 & .7753 \\
50 & .8210 & .7653 & .7927 & .7814 & .8036 & .7687 \\
60 & .8318 & .7700 & .7940 & .7882 & .8063 & .7813 \\
70 & .8223 & .7880 & .7873 & .7653 & .8083 & .7753 \\
\rowcolor{black!7}
80 & \textbf{.8358} & \textbf{.7920} & \textbf{.8053} & \textbf{.7902} & .8130 & \textbf{.7973} \\
90 & .8156 & .7820 & .7867 & .7801 & .8043 & .7873 \\
100 & .8149 & .7820 & .7960 & .7740 & .7936 & .7820 \\
\hline
\end{tabularx}

\end{table}

\subsection{Statistical Stability Across Training Runs}
\label{sec:d3_statistical_stability}
\autoref{tab:d3_stability} shows that \method{} maintains consistent performance across independent training runs on all three benchmarks. On BigCloneBench, F1 varies by only $.011$ standard deviation across seeds, with values ranging from $.732$ to $.762$, while accuracy remains within a narrow range of $.924$--$.933$. CodeNet shows similarly small variation, with F1 of $.736\pm.003$ and accuracy of $.678\pm.005$. AtCoder exhibits slightly larger variation than the other datasets, but the effect remains limited: F1 varies by $.013$ standard deviation (range $.791$--$.821$) and accuracy by $.018$ (range $.768$--$.812$). These results are consistent with the single-run scores reported in \autoref{sec:rq2}.

\begin{table}[!htbp]
\centering
\caption{Statistical stability of \method{} over five independent
training runs. Mean and standard deviation are computed across random seeds; the final column reports a 95\% Student-t confidence interval for the mean of the per-seed results.}
\label{tab:d3_stability}

\scriptsize
\setlength{\tabcolsep}{4.2pt}
\renewcommand{\arraystretch}{1.08}

\begin{tabular}{@{}llccc@{}}
\toprule
\textbf{Dataset}
& \textbf{Metric}
& \textbf{Mean $\pm$ Std}
& \textbf{Range}
& \textbf{95\% CI} \\
\midrule

BigCloneBench
& Prec.
& .724 $\pm$ .028
& [.687, .760]
& [.690, .759] \\

& Rec.
& .780 $\pm$ .031
& [.750, .828]
& [.742, .819] \\

& F1
& .750 $\pm$ .011
& [.732, .762]
& [.736, .765] \\

& Acc.
& .929 $\pm$ .004
& [.924, .933]
& [.924, .934] \\

\midrule
AtCoder
& Prec.
& .756 $\pm$ .025
& [.720, .783]
& [.725, .787] \\

& Rec.
& .865 $\pm$ .007
& [.859, .876]
& [.857, .874] \\

& F1
& .807 $\pm$ .013
& [.791, .821]
& [.790, .823] \\

& Acc.
& .793 $\pm$ .018
& [.768, .812]
& [.770, .815] \\

\midrule
CodeNet
& Prec.
& .624 $\pm$ .005
& [.620, .633]
& [.618, .631] \\

& Rec.
& .897 $\pm$ .010
& [.887, .914]
& [.884, .909] \\

& F1
& .736 $\pm$ .003
& [.733, .739]
& [.732, .740] \\

& Acc.
& .678 $\pm$ .005
& [.674, .686]
& [.672, .685] \\

\bottomrule
\end{tabular}
\end{table}

\subsection{What the Learned Latent Graph Captures}
\label{sec:d4_latent_graph_interpretation}
We examine the latent graphs produced by \method{}. As described in \autoref{sec:latent_induction}, the assignment matrix $P$ maps the input AST/DDG nodes of each code fragment to $m=32$ latent nodes through soft assignment, following the general idea of differentiable graph pooling \citep{ying2018diffpool}. For this analysis only, we group the 24 canonical node types of \autoref{sec:input_representation} into seven categories: \emph{Control}, \emph{Declaration}, \emph{Operator}, \emph{Call/Access}, \emph{Literal}, \emph{Structure/Meta}, and \emph{Unknown}. Let $\mathcal{C}_{\mathrm{val}}$ be the set of distinct validation fragments and let $\kappa(v)$ denote the category of input node $v$. For latent node $u$, we define its assignment profile as
\begin{equation}
Q_{uk}
=
\frac{
\sum_{c\in\mathcal{C}_{\mathrm{val}}}
\sum_{v:\kappa(v)=k} P_{vu}(c)
}{
\sum_{c\in\mathcal{C}_{\mathrm{val}}}
\sum_v P_{vu}(c)
},
\label{eq:d4_slot_profile}
\end{equation}
so each row of $Q$ gives the distribution of canonical-node groups assigned to a latent node. To examine the learned connections between these nodes, we also compute the mean weighted adjacency over the validation fragments, following the latent graph learning setting $\bar{A}=1/{|\mathcal{C}_{\mathrm{val}}|}\sum_{c\in\mathcal{C}_{\mathrm{val}}}A_{\Theta}(c)$~\citep{kipf2018nri,chen2020idgl,zhu2021gslsurvey}.

\begin{figure}[!htbp]
    \centering
    \includegraphics[
        width=1.0\columnwidth
    ]{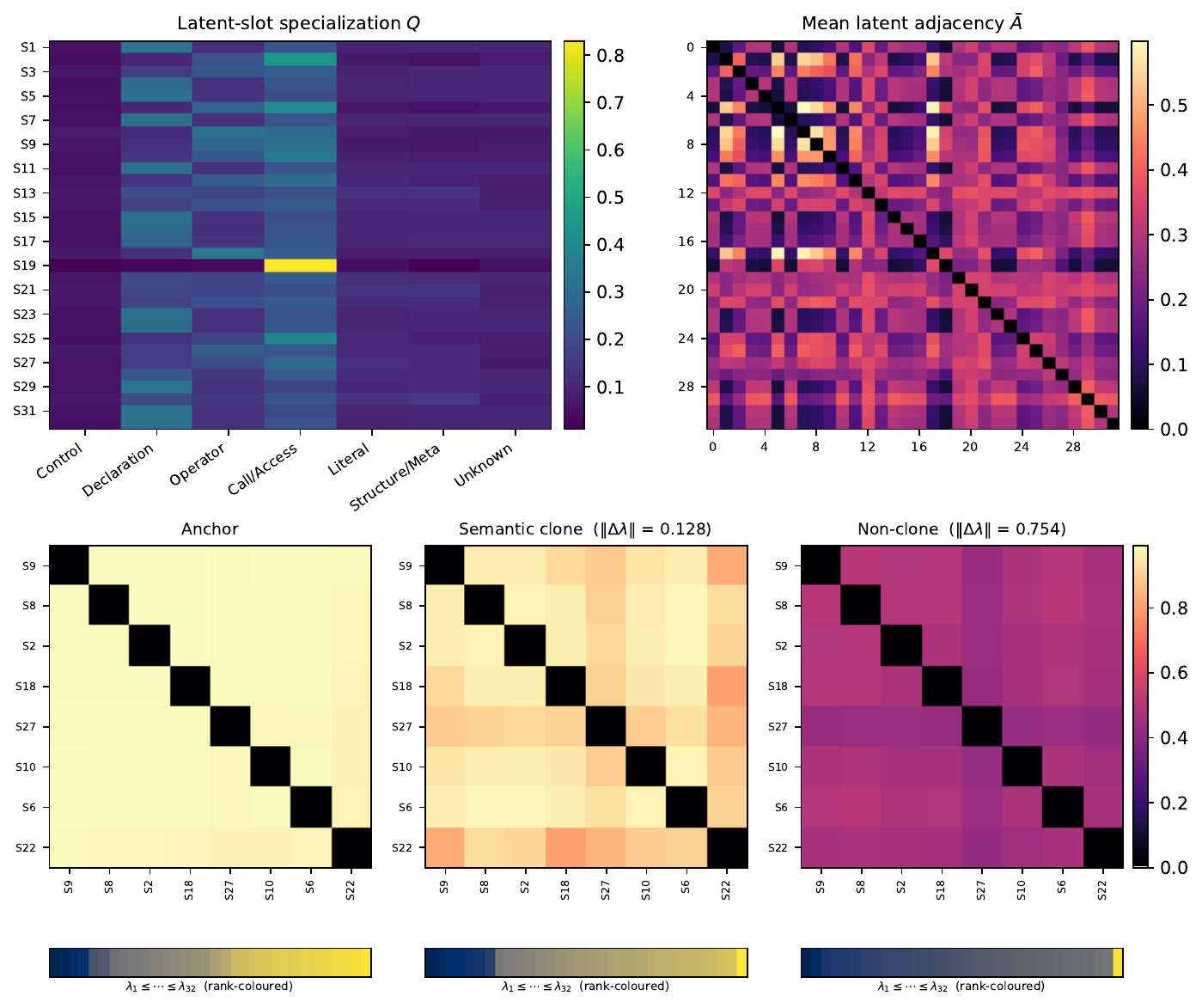}
    \caption{Interpretation of the learned latent graph. \emph{Left:} latent-slot specialization matrix $Q\in\mathbb{R}^{32\times7}$, where rows denote shared latent slots and columns denote coarse canonical node groups. \emph{Right:} validation-set mean latent adjacency $\bar{A}$ over the same latent slots. \emph{Bottom:} a representative anchor--clone--non-clone triplet with the corresponding normalized-Laplacian eigenvalue spectra. Only eight prominent latent slots are visualized for clarity. $|\Delta\Lambda|$ denotes the mean absolute difference between ordered eigenvalue vectors.}
    \label{fig:latent_graph_analysis}
\end{figure}

\autoref{fig:latent_graph_analysis} shows a correctly classified validation example. In the example, the clone differs from the anchor by $|\Delta \Lambda| = 0.328$ in eigenvalue distance, compared with $0.754$ for the non-clone. The assignment profile shows that some latent nodes favor particular canonical groups, while the mean adjacency shows that learned edge weights are not uniform across latent-node pairs.

\subsection{Computational Cost Analysis}
\label{sec:d5_computational_analysis}
\autoref{tab:d5_complexity_stats} shows that the training cost of \method{} is dominated by the backward pass, which accounts for $44.84\%$ of the per-batch runtime. Input processing, including data loading and host-to-device transfer, contributes another $32.18\%$, while the components introduced by the proposed representation remain lightweight. Latent graph induction accounts for only $5.79\%$ of the runtime, and spectral descriptors require $5.10\%$. Eigendecomposition, despite being the main additional operation introduced by the spectral representation, contributes only $1.07\%$ of the total batch cost. \method{} processes each batch of size $32$ in approximately $86.65$ ms, with a complete training epoch requiring about $41$ minutes on BigCloneBench, $24$ minutes on AtCoder, and $3$ minutes on CodeNet.

\begin{table}[!htbp]
\centering
\caption{Training computational cost of \method{}. Stage-level values report per-batch mean $\pm$ standard deviation, 95\% confidence interval, and runtime share. Aggregate rows report per-batch, per-input, and per-epoch costs (batch size is $32$).}
\label{tab:d5_complexity_stats}

\scriptsize
\setlength{\tabcolsep}{4.2pt}
\renewcommand{\arraystretch}{1.00}

\begin{adjustbox}{width=0.9\columnwidth}
\begin{tabular}{lccc}
\toprule
\textbf{Stage}
& \textbf{Mean $\pm$ Std. (ms)}
& \textbf{95\% CI (ms)}
& \textbf{Batch Share (\%)}
\\

\midrule
\rowcolor{black!7}
\multicolumn{4}{l}{\textit{Input}}
\\
\midrule
Data loading (wait)
& $16.48 \pm 2.66$
& $[16.46, 16.49]$
& $18.95 \pm 2.05$
\\
Host-to-device copy
& $11.42 \pm 0.38$
& $[11.42, 11.42]$
& $13.23 \pm 0.72$
\\

\midrule
\rowcolor{black!7}
\multicolumn{4}{l}{\textit{Encoder}}
\\
\midrule
Embedding + input norm
& $0.77 \pm 0.24$
& $[0.77, 0.77]$
& $0.89 \pm 0.05$
\\
Relational GNN layers
& $4.96 \pm 1.82$
& $[4.95, 4.97]$
& $5.75 \pm 0.36$
\\

\midrule
\rowcolor{black!7}
\multicolumn{4}{l}{\textit{Latent graph}}
\\
\midrule
Latent slot attention
& $4.40 \pm 1.18$
& $[4.39, 4.40]$
& $5.04 \pm 0.69$
\\
Latent adjacency
& $0.23 \pm 0.13$
& $[0.23, 0.23]$
& $0.26 \pm 0.06$
\\
Latent graph refinement
& $0.43 \pm 0.19$
& $[0.43, 0.43]$
& $0.49 \pm 0.18$
\\

\midrule
\rowcolor{black!7}
\multicolumn{4}{l}{\textit{Spectral}}
\\
\midrule
Spectral descriptors
& $4.43 \pm 1.19$
& $[4.42, 4.43]$
& $5.10 \pm 0.35$
\\
\quad of which: eigendecomposition
& $0.93 \pm 1.01$
& $[0.92, 0.93]$
& $1.07 \pm 0.07$
\\

\midrule
\rowcolor{black!7}
\multicolumn{4}{l}{\textit{Head}}
\\
\midrule
Pair comparison
& $1.75 \pm 0.20$
& $[1.75, 1.75]$
& $2.02 \pm 0.17$
\\
Classifier
& $0.52 \pm 0.47$
& $[0.52, 0.53]$
& $0.60 \pm 0.18$
\\

\midrule
\rowcolor{black!7}
\multicolumn{4}{l}{\textit{Backward}}
\\
\midrule
Loss
& $1.98 \pm 0.81$
& $[1.97, 1.98]$
& $2.27 \pm 0.30$
\\
Backward pass
& $38.80 \pm 3.19$
& $[38.79, 38.81]$
& $44.84 \pm 1.70$
\\
Optimizer step
& $0.49 \pm 1.06$
& $[0.49, 0.50]$
& $0.56 \pm 1.00$
\\

\midrule
\rowcolor{black!7}
\multicolumn{4}{l}{\textit{Aggregate cost}}
\\
\midrule
\textbf{Total per batch}
& \textbf{$86.65 \pm 9.46$}
& \textbf{$[86.60, 86.69]$}
& \textbf{$100$}
\\
Total per input
& $1.354 \pm 0.149$
& $[1.353, 1.355]$
& --
\\

\midrule
\rowcolor{black!7}
\multicolumn{4}{l}{\textit{Total per epoch}}
\\
\midrule
\textbf{BigCloneBench}
& \textbf{$41\mathrm{m}\,26\mathrm{s} \pm 1\mathrm{m}\,6\mathrm{s}$}
& --
& --
\\
\textbf{AtCoder}
& \textbf{$23\mathrm{m}\,43\mathrm{s} \pm 47\mathrm{s}$}
& --
& --
\\
\textbf{CodeNet}
& \textbf{$3\mathrm{m}\,12\mathrm{s} \pm 4\mathrm{s}$}
& --
& --
\\

\bottomrule
\end{tabular}
\end{adjustbox}

\end{table}

\subsection{Limitations and Future Work}
\label{sec:limitations}
The broad experimental study, especially the cross-language evaluation in \autoref{sec:rq3}, requires us to cap training at four epochs under our available computational budget. Therefore, the reported results do not represent fully converged models. Moreover, \method{} involves several design choices that affect the learned latent graph and its spectral representation, including graph induction parameters, spectral descriptor configuration, and optimization objectives. These choices are not exhaustively tuned due to the cost of repeated large-scale experiments. The current spectral representation is also derived from an initial graph containing only AST and projected DDG information, with input graphs limited to 256 nodes and compressed into a fixed number of latent slots. Finally, the latent topology is learned specifically for clone detection, and there is no explicit constraint that the resulting graph should preserve properties useful for other program analysis tasks.

Future work can investigate larger training budgets, systematic optimization of the latent graph and spectral representation parameters, and richer input representations that incorporate additional structural views. Beyond clone detection, an interesting direction is to study whether the learned latent graph provides a useful general-purpose representation for other software engineering tasks, such as vulnerability detection and code search. The latent graph induction process could also be adapted through transfer learning or alternative objectives, including self-supervised learning over large unlabeled repositories, to discover program structures that are not tied to a single downstream task.

\section{Threats to Validity}
\label{sec:validity}
We discuss potential threats that may affect the credibility and generalizability of our results, grouped into internal, external, and construct validity.

\subsection{Internal Validity}
Internal validity concerns whether the observed effects stem from our methods rather than uncontrolled factors. The four-epoch training cap may favor models that converge faster, although \autoref{sec:num_epochs} suggests this works against \method{}, but we cannot rule out that it affects individual baselines differently. In addition, single-run results may be affected by training randomness; we assess this variability through the multi-seed analysis in \autoref{sec:d3_statistical_stability}.

\subsection{External Validity}
External validity concerns the generalizability of our findings. AtCoder and CodeNet are derived from programming contest problems, and their distributions may not reflect general software development. Moreover, our evaluation mostly covers function-level fragments, so the observed performance may not fully represent industrial codebases with different characteristics, such as larger systems, evolving projects, and collaborative development patterns.

\subsection{Construct Validity}
Construct validity concerns whether the evaluation setup measures the properties that the study aims to investigate. The meaning of a positive pair may vary across datasets, with BigCloneBench relying on human-assisted functional clone annotations, while AtCoder and CodeNet define clones based on shared problem identity. Therefore, benchmark labels should be interpreted as proxies for semantic equivalence rather than perfect indicators of functional similarity.

\section{Conclusions}
\label{sec:conclusion}
We presented \method{}, a Siamese latent graph learning network that induces a fixed-size latent graph from the AST and data-dependence information of each code fragment and derives a multi-scale spectral representation from its learned topology. Unlike previous approaches that extract features from fixed program graphs (AST, CFG, DDG, PDG, and CPG), \method{} lets clone supervision determine graph latent space that best captures functional similarity between codes through its spectral representation.

Our experiments support three main results. First, learning the graph structure before spectral analysis is decisive for obtaining discriminative code spectra. With an identical SNN classifier, the learned spectrum of \method{} improves F1 from .37 to .67 on BigCloneBench (+.30) and accuracy from .60 to .71 on AtCoder (+.11) compared with spectra extracted from fixed program graphs. Second, the complete model is competitive with established semantic clone detectors, obtaining .76 F1 on BigCloneBench, .82 F1 on AtCoder, and .69 accuracy on the four-language CodeNet benchmark, where it surpasses ASTNN and GraphCodeBert using only four epochs. Third, the learned representation transfers beyond test language pairs $Y$--$Y$. Across 60 unseen bridge-assisted transfer paths, \method{} exhibits the smallest degradation under language shift, decreasing by only .058 compared with .112, .162, .170, and .228 for GraphCodeBERT, ASTNN, DeepSim, and RtvNN, and adding intermediate-language supervision further improves transfer accuracy by up to .036.

\section{Declarations}




\subsection{Data Availability Statement}
The datasets generated and/or analyzed during the current study are available at:
\begin{itemize}
    \item \textbf{Data repository:}
    \begin{itemize}
        \item CodeNet: \url{https://www.kaggle.com/datasets/koushamoeini/codenet-4l}
        \item AtCoder: \url{https://www.kaggle.com/datasets/koushamoeini/atcoder}\\
        \item BigCloneBench: \url{https://www.kaggle.com/datasets/koushamoeini/codexglue}
    \end{itemize}
    \item \textbf{Code repository:} \url{https://github.com/Knowledge4Software/Spectral-Software}
\end{itemize}



\IfFileExists{cas-refs.bib}{%
  \bibliographystyle{elsarticle-harv}%
  \bibliography{cas-refs}%
}{%
  \typeout{WARNING: cas-refs.bib is missing; bibliography omitted.}%
}
\end{document}